\documentclass[12pt]{article}
\pdfoutput=1
\usepackage{jheppub}
\usepackage{amsmath}
\usepackage{amsfonts}
\usepackage{amssymb}
\usepackage{graphicx}

\usepackage[export]{adjustbox}
\newcommand{\bmat}{\left(\begin{array}}
\newcommand{\emat}{\end{array}\right)}

\def\yzero{\smash{\hbox{$y\kern-4pt\raise1pt\hbox{${}^\circ$}$}}}

\def\beq{\begin{equation}}
\def\eeq{\end{equation}}
\def\beqa{\begin{eqnarray}}
\def\eeqa{\end{eqnarray}}

\def\-{\hphantom{-}}

\def\s2{\frac{1}{\sqrt2}}

\def\beq{\begin{equation}}
\def\eeq{\end{equation}}
\def\beqa{\begin{eqnarray}}
\def\eeqa{\end{eqnarray}}

\def\IF{\relax{\rm I\kern-.18em F}}
\def\II{\relax{\rm I\kern-.18em I}}

\def\Dsl{\,\raise.15ex\hbox{/}\mkern-13.5mu D} 
\def\id{{\bf {1}}}

\def\IC{{\bf{C}}}
\def\IS{{\bf {S}}}
\def\IR{{\bf {R}}}
\def\IZ{{\bf {Z}}}
\def\IX{{\bf {X}}}

\def\IT{{\bf {T}}}

\def\NN{{\cal {N}}}

\newcommand{\drawsquare}[2]{\hbox{%
\rule{#2pt}{#1pt}\hskip-#2pt
\rule{#1pt}{#2pt}\hskip-#1pt
\rule[#1pt]{#1pt}{#2pt}}\rule[#1pt]{#2pt}{#2pt}\hskip-#2pt
\rule{#2pt}{#1pt}}

\newcommand{\fund}{~\raisebox{-.5pt}{\drawsquare{6.5}{0.4}}~}
\newcommand{\antifund}{~\overline{\raisebox{-.5pt}{\drawsquare{6.5}{0.4}}}~}

\catcode`\@=11   
\newdimen\@rotdimen
\newbox\@rotbox  

\def\@vspec#1{\special{ps:#1}}
\def\@rotstart#1{\@vspec{gsave currentpoint currentpoint translate
   #1 neg exch neg exch translate}}
\def\@rotfinish{\@vspec{currentpoint grestore moveto}}
\def\@rotr#1{\@rotdimen=\ht#1\advance\@rotdimen by\dp#1%
   \hbox to\@rotdimen{\hskip\ht#1\vbox to\wd#1{\@rotstart{90 rotate}%
   \box#1\vss}\hss}\@rotfinish}
\def\@rotl#1{\@rotdimen=\ht#1\advance\@rotdimen by\dp#1%
   \hbox to\@rotdimen{\vbox to\wd#1{\vskip\wd#1\@rotstart{270 rotate}%
   \box#1\vss}\hss}\@rotfinish}%
\def\@rotu#1{\@rotdimen=\ht#1\advance\@rotdimen by\dp#1%
   \hbox to\wd#1{\hskip\wd#1\vbox to\@rotdimen{\vskip\@rotdimen
   \@rotstart{-1 dup scale}\box#1\vss}\hss}\@rotfinish}%
\def\@rotf#1{\hbox to\wd#1{\hskip\wd#1\@rotstart{-1 1 scale}%
   \box#1\hss}\@rotfinish}%
\def\rotate{\@ifnextchar[{\@rotate}{\@rotate[l]}}
\def\@rotate[#1]#2{\setbox\@rotbox=\hbox{#2}\@nameuse{@rot#1}\@rotbox}

\catcode`\@=12

\begin{document}

\makeatletter
\@addtoreset{equation}{section}
\makeatother
\renewcommand{\theequation}{\thesection.\arabic{equation}}
\pagestyle{empty}
\vspace*{0.5in}
\vspace{1cm}
\begin{center}
\Large{\bf Aspects of Dynamical Cobordism in AdS/CFT
}
\\[8mm] 

\large{Jesús Huertas, Angel M. Uranga\\[4mm]}
\footnotesize{Instituto de F\'{\i}sica Te\'orica IFT-UAM/CSIC,\\[-0.3em] 
C/ Nicol\'as Cabrera 13-15, 
Campus de Cantoblanco, 28049 Madrid, Spain}\\ 
\footnotesize{\href{j.huertas@csic.es}{j.huertas@csic.es},  \href{mailto:angel.uranga@csic.es}{angel.uranga@csic.es}}

\vspace*{10mm}

\small{\bf Abstract} \\
\end{center}
\begin{center}
\begin{minipage}[h]{\textwidth}
The cobordism conjecture implies that consistent theories of Quantum Gravity must admit the introduction of boundaries.
We study the dynamical realization of the cobordism conjecture in type IIB in AdS$_5\times\IS^5$, 
using the existing gravity duals of 4d $\NN=4$ SYM with Gaiotto-Witten superconformal boundary conditions (near-horizon limits of D3-branes ending on NS5- and D5-branes).
We show that these configurations are, from the 5d perspective, dynamical cobordism solutions which start from an asymptotic AdS$_5$ vacuum and evolve until they hit an end of the world (ETW) brane with AdS$_4$ worldvolume. The latter displays localization of gravity, and provide a completion of the Karch-Randall (KR) AdS branes, in which the backreaction of running scalars replace the KR cusp in the warp factor with a smooth bump. The dynamical scalars are either in the $SO(6)$ invariant AdS$_5$ bulk sector (e.g. describing the $\IS^5$ size and its shrinking at the cobordism boundary) or brane localized (e.g. the $SO(6)\to SO(3)\times SO(3)$ squashing due to boundary conditions).
We introduce a novel double scaling limit which zooms into the ETW brane and makes localization of gravity manifest, and which shows a tantalizing relation with wedge holography.
We extend the picture to AdS$_5$ theories with less (super)symmetry, via  orbifolds and S-folds, leading to dynamical cobordisms for gravity duals of 4d theories with $\NN=2$ and $\NN=3$ supersymmetry.
\end{minipage}
\end{center}
\newpage
\setcounter{page}{1}
\pagestyle{plain}
\renewcommand{\thefootnote}{\arabic{footnote}}
\setcounter{footnote}{0}

\tableofcontents

\vspace*{1cm}

\newpage

\section{Introduction}

The cobordism conjecture \cite{McNamara:2019rup} implies that consistent theories of quantum gravity, in particular string theory, should admit the introduction of boundaries, configurations (possibly dressed with localized objects to absorb charges/fluxes) ending spacetime. We will refer to such configurations End of the World (ETW) branes. At the topological level, this has led to non-trivial results regarding anomalies and the existence of (possibly non-supersymmetric) extended states in string theory \cite{GarciaEtxebarria:2020xsr,Ooguri:2020sua,Montero:2020icj,Dierigl:2020lai,Hamada:2021bbz,Blumenhagen:2021nmi,Andriot:2022mri,Dierigl:2022reg,Debray:2023yrs}. At the level of physical realizations, it has led to the notion of dynamical cobordisms \cite{Buratti:2021yia,Buratti:2021fiv,Angius:2022aeq,Blumenhagen:2022mqw,Blumenhagen:2023abk}  (see also 
\cite{Dudas:2000ff,Blumenhagen:2000dc,Dudas:2002dg,Dudas:2004nd,Hellerman:2006nx,Hellerman:2006ff,Hellerman:2007fc} for related early work and \cite{Basile:2018irz,Antonelli:2019nar,Mininno:2020sdb,Basile:2020xwi,Basile:2021mkd,Mourad:2022loy,Angius:2022mgh,Angius:2023xtu} for other related recent developments), spacetime dependent solutions developing real codimension-1 singularities ending spacetime, at which certain scalars run off to infinite distance in field space \cite{Buratti:2021fiv}. Hence, dynamical cobordisms are useful probes of infinite distance limits in field space and the swampland distance conjecture \cite{Ooguri:2006in} (and in fact relate to other probes, such as small black holes \cite{Hamada:2021yxy,Angius:2022aeq,Angius:2023xtu} or 4d EFT strings \cite{Lanza:2020qmt,Lanza:2021qsu,Marchesano:2022avb}.

It is natural to explore the cobordism conjecture in the arguably best understood vacua in string theory, those arising in AdS/CFT holography, such as type IIB on AdS$_5\times\IS^5$ with $N$ units of RR 5-form flux. The simplest proposal for a boundary configuration for the latter would seem to simply correspond to shrinking the $\IS^5$ at some value of the radial coordinate in a Minkowski slicing of AdS$_5$, including $N$ explicit D3-branes to remove the flux; however, the backreaction of this large number of D3-branes would reconstruct a non-compact AdS$_5$ throat, preventing an actual end of spacetime. 

This puzzle can however be overcome by the introduction of a different class of boundaries, familiar in the setup known as AdS/BCFT or double holography \cite{Karch:2000gx,Takayanagi:2011zk}. The key idea is to consider the boundary of an AdS$_{d+1}$ spacetime\footnote{Our convention is to use the term {\em ETW boundary}, or just {\em boundary}, for the cobordism one, and {\em holographic} boundary for the conformal boundary at infinity.} to be localized at a codimension-1 AdS$_d$ slice. These so called Karch-Randall (KR) branes \cite{Karch:2000ct,Karch:2001cw,Karch:2000gx} have the property of localizing gravity on their worldvolume, and reaching off to the holographic boundary, so that they define a boundary in the dual CFT$_d$. Double holography is the statement that the gravitational theory on the AdS$_d$ worldvolume of the ETW brane is the gravity dual of a $(d-1)$-dimensional boundary CFT (BCFT) of the $d$-dimensional CFT holographically dual to the AdS$_{d+1}$ gravitational bulk. This kind of AdS ETW branes in AdS$_3$/CFT$_2$ also play a prominent role in the recent explanation of the black hole Page curve and quantum islands \cite{Almheiri:2019hni,Almheiri:2019psy, Chen:2020uac,Chen:2020hmv,Geng:2020fxl,Geng:2021mic,Uhlemann:2021nhu,Demulder:2022aij}.

Despite their importance, most explorations of double holography and quantum islands involve a bottom-up simplified modeling of ETW branes as sharp boundaries, with no microscopic description. However, for AdS$_5\times\IS^5$ there is a class of 10d supergravity solutions, including D5- and NS5-branes, which provide such description for boundaries preserving an $SO(3)\times SO(3)$ subgroup of the $SO(6)$ symmetry \cite{DHoker:2007zhm,DHoker:2007hhe,Aharony:2011yc,Assel:2011xz,Bachas:2017rch,Bachas:2018zmb} (see also \cite{Raamsdonk:2020tin,VanRaamsdonk:2021duo,Demulder:2022aij,Karch:2022rvr} for recent applications, and \cite{DHoker:2008lup,DHoker:2008rje,DHoker:2009lky,DHoker:2009wlx} for related results on AdS$_4\times\IS^7$ and AdS$_7\times \IS^4$); these configurations are dual to 4d $\NN=4$ SYM with a boundary, with the $OSp(4|4)$ superconformal symmetry preserving boundary conditions studied in \cite{Gaiotto:2008sa, Gaiotto:2008ak}. The 5-branes source NSNS and RR 3-form fluxes, whose Chern-Simons couplings remove the 5-form flux and allow the $\IS^5$ to shrink  (see \cite{Buratti:2021yia} for a similar phenomenon in the Klebanov-Strassler solution \cite{Klebanov:2000hb}) with no non-compact throats arising from the backreaction.

In this paper we discuss aspects of this class of solutions, with emphasis in extracting their interpretation from the perspective of the 5d theory, and their relation with the bottom-up approach. We show that the 5d description takes the form of a dynamical cobordism with fields running along the coordinate transverse to the AdS$_4$ slicing. The description includes the 5d metric and the dilaton, but a prominent role is played by the scalar breathing mode of the $\IS^5$. In particular, it is necessary to define the shrinking of the $\IS^5$ at the boundary, in a way dictated by the local ETW brane description in \cite{Angius:2022aeq}. In addition, it smooths out the cusp of the naive 5d KR metric into a bump, which nevertheless retains the feature of localizing gravity. Our analysis allows to systematically include further KK modes, and enrich the picture. For instance, they encode the breaking $SO(6)\to SO(3)\times SO(3)$ via scalars localized on the ETW brane. We also introduce a double scaling limit of the solutions (interestingly related to configurations in \cite{Assel:2011xz,Bachas:2017rch} and to wedge holography \cite{Akal:2020wfl,VanRaamsdonk:2021duo}), which allows to isolate the ETW brane dynamics and make some of its properties manifest. We carry out the discussion mainly for AdS$_5\times \IS^5$, but subsequently discuss S-fold and orbifold quotients preserving the structure of the 10d solution, hence providing a description for ETW branes in AdS$_5$ spaces in cases with lower supersymmetry, gravity duals of 4d $\NN=3$ and $\NN=2$ field theories with boundaries.

This paper is organized as follows. In Section \ref{sec:overview} we review the 10d solutions describing AdS$_5\times\IS^5$ ending on ETW brane configurations and their interpretation as gravity duals of 4d $\NN=4$ SYM with boundaries defined by NS5- and D5-branes. Section \ref{sec:bagpipes} describes the general class of solutions, section \ref{sec:case-of-one} particularizes to the ETW branes of AdS$_5\times\IS^5$, and section \ref{sec:holobound} discusses the holographic relation. In Section \ref{sec:cobordism} we interpret them as dynamical cobordisms. Section \ref{sec:bagpipe-cobordism} describes the solutions as dynamical cobordisms of several AdS$_5\times\IS^5$ asymptotic regions, or cobordisms to nothing for a single one. In section \ref{sec:eft} we provide a 5d interpretation of the solutions after reduction on the $\IS^5$, including the 5d metric and localization of gravity, and  bulk and localized scalars. Section \ref{sec:double-scaling} discusses a double scaling limit isolating the dynamics of the ETW brane. In section \ref{sec:scaling} we present a scaling symmetry of the solutions, and section \ref{sec:zoom} describes the scaling limit, for which we provide an alternative interpretation in section \ref{sec:revisit}; section \ref{sec:reduct-limit} deals with the effective theory reduction of the resulting solutions. In Section \ref{sec:n2} we extend the discussion to gravity duals of 4d $\NN=2$ orbifold field theories. Section \ref{sec:quiver} introduces the field theories and their gravity duals. In section \ref{sec:etw-n2} we present the orbifold ETW configurations, which we recast in section \ref{sec:branebox}  in terms of T-dual brane box configurations. Section \ref{sec:sfolds} discusses ETW configurations in gravity duals of 4d S-fold field theories. In section \ref{sec:quiverfold} we introduce the field theories, and in section \ref{sec:etw-sfold} we describe their gravity duals and the corresponding solutions for ETW configurations. In Section \ref{sec:conclusions} we offer some final remarks.

{\bf Note added:} As this paper was being preparing for submission we noticed \cite{DeLuca:2023kjj}, which deals with gravity localization in this and other setups.

\section{Overview of ETW branes for AdS$_5\times \IS^5$}
\label{sec:overview}

In this section we review the  configurations ending AdS$_5\times \IS^5$ on ETW configurations along an AdS$_4$ slice, on which the $\IS^5$ is able to contract while suitable NSNS and RR 3-form fluxes manage to remove the $N$ units of RR 5-form flux on it. The ETW configuration reaches off to the holographic boundary, hence is holographically dual to a boundary of the 4d $\NN=4$ SYM theory. The geometries, and their holographic interpretation in BCFT, have been studied in \cite{DHoker:2007zhm,DHoker:2007hhe,Aharony:2011yc,Assel:2011xz,Bachas:2017rch,Bachas:2018zmb}, see also \cite{Raamsdonk:2020tin,VanRaamsdonk:2021duo,Demulder:2022aij} for recent applications. We merely outline the results useful for our purposes, and refer the reader for further details. For concreteness, we follow the notation and conventions in \cite{Aharony:2011yc}.

\subsection{Bagpipe geometries}
\label{sec:bagpipes}

The symmetry of the solutions is the 3d $\NN=4$ superconformal group $OSp(4|4)$, whose bosonic symmetry is $SO(2,3)\times SO(3)\times SO(3)$. They have the structure
\beqa
{\rm AdS}_4\times \IS_1^2\times \IS_2^2\times \Sigma\, ,
\eeqa
where $\Sigma$ is an oriented Riemann surface, over which the AdS$_4$ and the two $\IS^2$'s vary. The ansatz for the Einstein frame 10d metric is
\beqa
ds^2= f_4^2 ds^2_{AdS_4}+f_1^2 ds^2_{\IS_1^2}+f_2^2 ds^2_{\IS_2^2}+ds^2_\Sigma\, .
\label{ansatz}
\eeqa
Here $f_1$, $f_2$, $f_3$ are functions of a complex coordinate $w$ of $\Sigma$, in terms of which 
\beqa
ds^2_\Sigma=4\rho^2 |dw|^2\, ,
\label{2dmetric}
\eeqa
for some real function $\rho$.
There are also non-trivial backgrounds for the NSNS and RR 2-forms and the RR 4-form, which we will mostly skip in our discussing, refering the reader to the references for details.

There are closed expressions for the different functions in the above metric, describing the BPS solution. As an intermediate step, we define the real functions
\beqa
& W\equiv \partial_w h_1\partial_{\bar w}h_2+\partial_w h_2\partial_{\bar w}h_1 
\; \;,\;\; N_1\equiv 2h_1h_2|\partial_w h_1|^2-h_1^2 W \;\; ,\;\;N_2\equiv 2h_1h_2|\partial_w h_2|^2-h_2^2 W\, .
\nonumber\\
\label{wnn}
\eeqa
The dilaton is given by
\beqa
e^{2\Phi}=\frac{N_2}{N_1}
\label{dilaton}
\eeqa
and the functions are given by
\beqa
\rho^2=e^{-\frac 12 \Phi}\frac{\sqrt{N_2|W|}}{h_1h_2}\; ,\; f_1^2=2e^{\frac 12\Phi} h_1^2\sqrt{\frac{|W|}{N_1}}\; ,\; f_2^2=2e^{-\frac 12\Phi} h_2^2\sqrt{\frac{|W|}{N_2}}\; ,\; f_4^2=2e^{-\frac 12\Phi} \sqrt{\frac{N_2}{|W|}}\nonumber\\
\label{the-fs}
\eeqa
The functions $h_1$, $h_2$ are defined using an auxiliary genus $g$ hyper-elliptic Riemann surface, parametrized by a coordinate $u$ in the lower half-plane ${\rm Im}\, u\leq 0$. This coordinate will be related to the variable $w$ above later on, so we denote the Riemann surface by $\Sigma$. It  is defined by the hyper-elliptic equation
\beqa
s^2(u)=(u-e_1)\prod_{k=1}^g(u-e_{2k})(u-e_{2k+1})\quad {\rm with}\, e_i\in \IR\, .
\eeqa
Note that we have fixed the $SL(2,\IR)$ symmetry to locate the branch point $e_{2g+2}$ at infinity.
The holomorphic differentials $\partial h_1$, $\partial h_2$ on this surface (i.e. 1-forms with branch points at $e_i$) are defined as
\beqa
\partial h_1=-i\frac{P(u)}{Q_1(u)}{s^3(u)}du\quad , \quad\partial h_2=-\frac{P(u)}{Q_2(u)}{s^3(u)}du
\eeqa
where $P$, $Q_1$, $Q_2$ are real polynomials of degrees $2g$, $g+1$, $g+1$, respectively, and the following structure of zeros 
\beqa
P(u)=\prod_{a=1}^g(u-u_a)(u-{\bar u}_a)\quad,\quad Q_1(u)=\prod_{b=1}^{g+1}(u-\alpha_b) \quad ,\quad
Q_2(u)=\prod_{b=1}^{g+1}(u-\beta_b)
\eeqa
where ${\rm Im}(u_a)\leq 0$. Regularity of the solution requires the following ordering of the zeroes
\beqa
\alpha_{g+1}\leq e_{2g+1}\leq \beta_{g+1}\leq e_{2g} \leq\ldots \leq \alpha_b\leq e_{2b-1}\leq \beta_b\leq \ldots \leq e_2\leq \alpha_1\leq e_1\leq \beta_1 \quad\quad
\label{general-order}
\eeqa
namely, alternating $\alpha$'s and $\beta$'s, with branch points $e_i$ separating them. The parameters $u_a$ can be determined in terms of these and are hence not independent quantities. The generic solution where the ordering (\ref{general-order}) is obeyed with strict inequalities corresponds to regular solutions; when some of the values coincide, the solution develops interesting singularities, see later.

\begin{figure}[htb]
\begin{center}
\includegraphics[scale=.35]{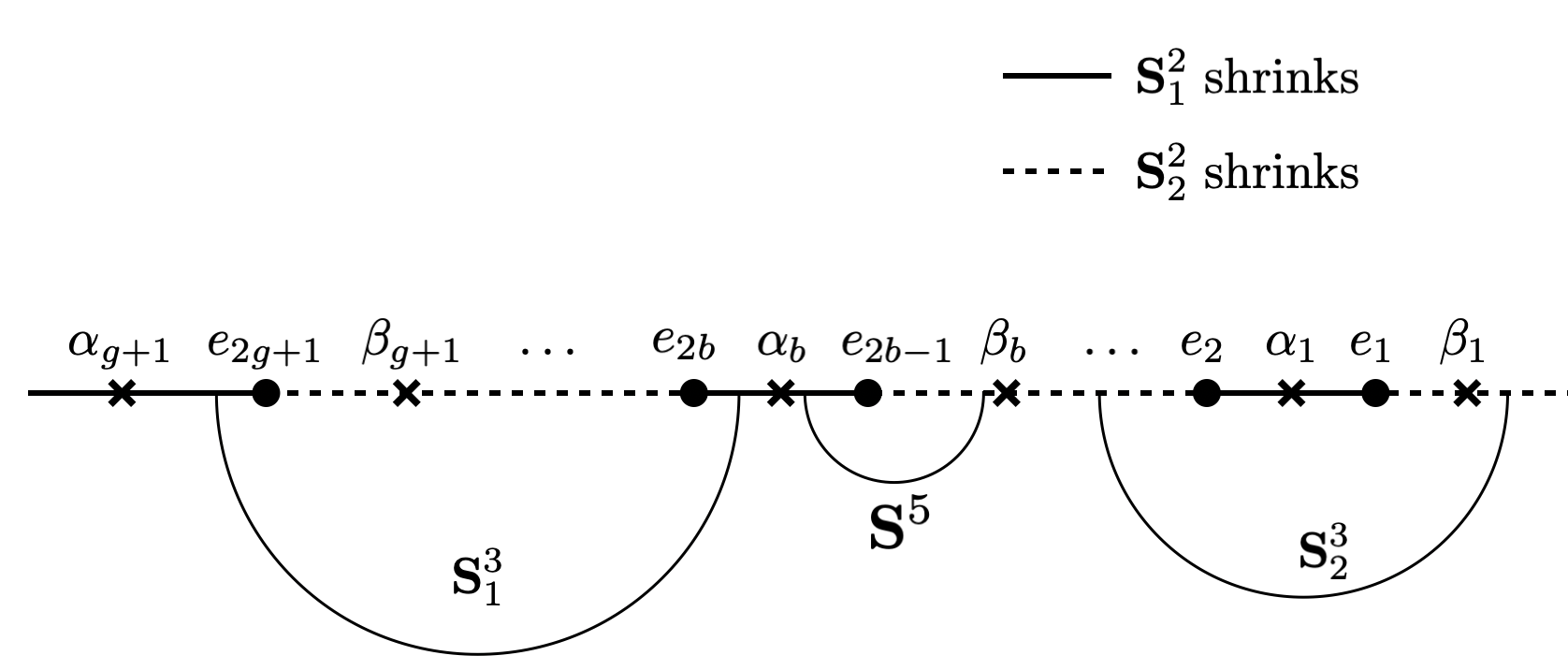}
\caption{\small Fibration of $\IS_1^2\times\IS_2^2$ over the $u$-plane, indicating the degenerations of fibres over the real line, and the kinds of non-trivial cycles in the geometry.}
\label{fig:sigma}
\end{center}
\end{figure}

The distribution of branch points on the real line ${\rm Im}(u)=0$ encodes interesting properties of the topology of the solution, see Figure \ref{fig:sigma}. The fibration of $\IS_1^2$ and $\IS_2^2$ over this real line has the following structure: the $\IS_1^2$ shrinks to zero size on top of the segments $(e_{2p},e_{2p-1})$, for $p=1,\ldots, g+1$ (and $e_{2g+2}=-\infty$), while the $\IS_2^2$ shrinks to zero size on top of the segments $(e_{2p-1},e_{2p-2})$, for $p=1,\ldots, g+1$ (and $e_0\equiv +\infty$). This implies that the boundary of the Riemann surface is not a boundary of the full geometry, as the closing off of the two-spheres seals it off.

Hence, the fibration of $\IS_1^2\times \IS_2^2$ over a segment  in the lower half-plane, with its two endpoints on this real line and separated by one brach point, has different two-spheres shrinking at its endpoints, so it has the topology of an $\IS^5$. Actually, at these points the 10d geometry develops a semi-infinite spike, corresponding to an AdS$_5\times \IS^5$ region. The values of the RR 5-form flux $N$ and $g_s$ will be discussed later on. 

\begin{figure}[htb]
\begin{center}
\includegraphics[scale=.35]{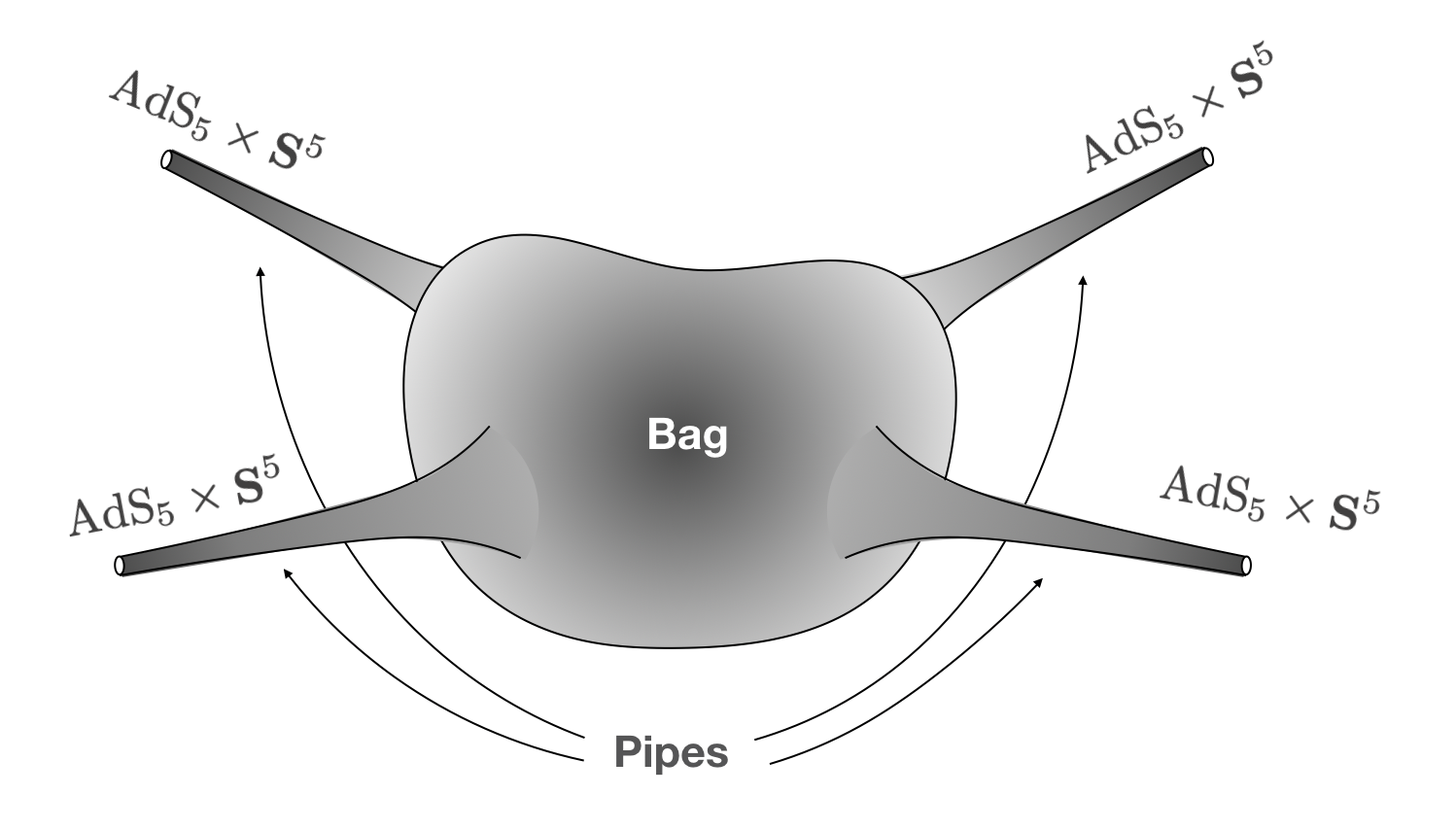}
\caption{\small The bagpipe geometries.
}
\label{fig:bagpipe}
\end{center}
\end{figure}

These geometries were dubbed `bagpipes'  in \cite{Bachas:2018zmb}, as they morally correspond to AdS$_4\times \IX_6$ (the bag), with $\IX_6$ a compact 6d manifold, with AdS$_5\times \IS^5$ asymptotic regions sticking out of it (the pipes), see Figure \ref{fig:bagpipe}.

On the other hand, fibering $\IS_1^2$ and $\IS_2^2$ over a segment  in the lower half-plane, with its two endpoints on this real line and separated by two branch points, has the same two-sphere collapsing at its endpoints, so it has the topology of $\IS^2\times\IS^3$. The $\IS^3$ acquires a sharp physical meaning if one considers the limit in which the two branch points coincide (and so does the corresponding $\alpha$ or $\beta$ between them in the ordering (\ref{general-order}). If we have the collapse of a subset $(e,\alpha,e)$, the resulting geometry locally corresponds to a NS5-brane solution; on the other hand, the collapse of a subset $(e,\beta,e)$, the resulting geometry locally corresponds to a D5-brane solution. The interpretation of these objects will become clear in the holographic description in  section \ref{sec:holobound}. 

One may also consider mixed collapses $(\alpha, e, \beta)$ or $(\beta,e,\alpha)$. Interestingly, the effect is the vanishing of the value of the RR 5-form flux in the would-be local AdS$_5\times \IS^5$ spike, meaning that the $\IS^5$ is actually contractible and the spike is closed off into a smooth point of the geometry.

\subsection{Towards one AdS$_5\times \IS^5$ asymptotic region}
\label{sec:case-of-one}

Note that the collapse of branch points leads to an effective reduction of the genus of the Riemann surface, hence of the number of asymptotic AdS$_5\times \IS^5$ regions. In the following we focus on the case of at most two asymptotic AdS$_5\times \IS^5$ regions. This arises when the ordering (\ref{general-order}) collapses as
\beqa
\alpha \; \underbrace{(e\;\beta\;e)\; \alpha \; (e\;\beta\;e)\; \alpha\; (e\;\beta\;e)}_{m}\; \alpha\; e\; \beta \; \underbrace{(e\;\alpha\;e)\;\beta \; (e\;\alpha\;e)\;\beta \; (e\;\alpha\;e)}_n\;\beta\, ,
\label{nsd-order}
\eeqa
with $n+m=g$.
This corresponds to a configuration with two asymptotic AdS$_5\times \IS^5$ regions, around the isolated branch point $e$ in the middle, and that at infinity. In addition, there are $n$ NS5-brane positions and $m$ D5-brane positions. In the following we focus on these configurations. Let us introduce the quantities $k_a^2$, $l_b^2$ with $a=1,\ldots, n$ and $b=1,\ldots, m$ to define the positions of the NS5- and D5-branes, respectively, so that the ordering is
\beqa
\alpha_{g+1}<-l_m^2<\alpha_g<\cdots<\alpha_{n+2}<-l_1^2<\alpha_{n+1}<0<\beta_{n+1}<k_n^2<\cdots<\beta_2<k_1^2<\beta_1\nonumber
\, ,\\
\label{the-order}
\eeqa
where, without loss of generality, we choose to locate the middle branch point at $u=0$.

The interpretation of these geometries is the following. For large $|u|$, the geometry asymptotes to AdS$_5\times\IS^5$ with some RR 5-form flux.  For intermediate values of $|u|$, there is a non-trivial warping, describing some transition region. Finally, the solution near the origin $u=0$ describes AdS$_5\times\IS^5$ with some (possibly different) RR 5-form flux. The solution thus describes an effective domain wall configuration connecting two AdS$_5\times\IS^5$, described by a `fat brane' with AdS$_4$ symmetry. Hence it corresponds to a string theory version of the Karch-Randall branes \cite{Karch:2000ct} (already anticipated in \cite{Karch:2001cw}).

\medskip

We focus on the case in which, in a subsequent step, we also collapse of the middle cluster by setting $\alpha_{n+1}=\beta_{n+1}=0$, namely (\ref{the-order}) turns into
\beqa
\alpha_{g+1}<-l_m^2<\alpha_g<\cdots<\alpha_{n+2}<-l_1^2<0<k_n^2<\cdots<\beta_2<k_1^2<\beta_1\nonumber
\, ,\\
\label{the-order-bis}
\eeqa
As explained above, this corresponds to closing off the asymptotic AdS$_5\times \IS^5$ spike at $u=0$, leaving only that at infinity\footnote{The inquisitive reader may wonder if this last asymptotic region can be closed off as well. This is in fact the punchline of section \ref{sec:double-scaling}.}.

\begin{figure}[htb]
\begin{center}
\includegraphics[scale=.3]{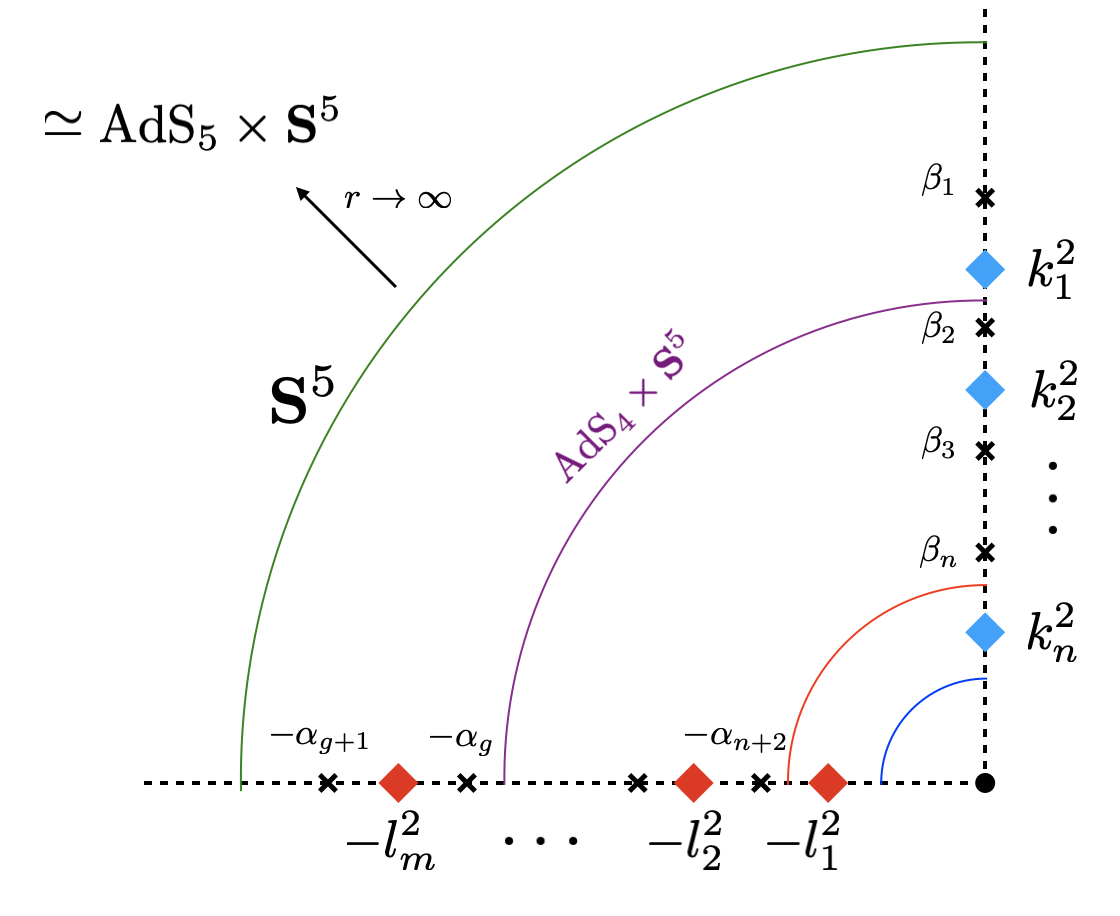}
\caption{\small Fibration of $\IS_1^2\times\IS_2^2$ over the $w$-plane for one asymptotic AdS$_5\times\IS^5$ region. The arcs correspond to slices with AdS$_4\times \IS^5$ geometry. We indicate the location of NS5- and D5-branes in the ETW region and the asymptotic AdS$_5\times\IS^5$. (Inspired in Fig. 4 of \cite{Raamsdonk:2020tin}).}
\label{fig:fivebranes}
\end{center}
\end{figure}

It is useful to introduce the coordinate change $u=w^2$. This means that $w$ parametrizes the second quadrant in the complex plane, and that the boundary of the Riemann surfaces contains two semi-infinite lines: the negative real axis (along with the NS5-brane positions are located) and the positive imaginary axis (along which the D5-brane positions are located). The closed-off asymptotic AdS$_5\times \IS^5$ sits at $w=0$, while the only left-over asymptotic AdS$_5\times \IS^5$ is at infinity. This is shown in Figure \ref{fig:fivebranes}.

The resulting structure of the functions $h_1$, $h_2$ is
\beqa
&&h_1=\;4{\rm Im}(w)+2\sum_{b=1}^m{\tilde d}_b \log\left(\frac{|w+il_b|^2}{|w-il_b|^2}\right)\nonumber\\
&&h_2=-4{\rm Re}(w)-2\sum_{a=1}^n d_a \log\left(\frac{|w+k_a|^2}{|w-k_a|^2}\right)\, ,
\label{the-hs}
\eeqa
where
\beqa
d_a\equiv \frac{(\beta_a-k_a^2)}{2k_a}\prod_{c\neq a}^n\frac{(k_a^2-\beta_c)}{(k_a^2-k_c^2)}\quad,\quad
{\tilde d}_b\equiv \frac{(-\alpha_{b+n+1-l_b^2}-l_b^2)}{2l_b}\prod_{c\neq b}^m\frac{(l_b^2+\alpha_{c+n+1})}{(l_b^2-l_c^2)}\, .
\label{the-ds}
\eeqa
We note that the parameters $d_a$, ${\tilde d}_b$ are positive. Moreover, the computation of the NSNS and RR 3-form fluxes on the corresponding $\IS^3$'s shows that they are related to the numbers $n_a$, $m_b$ of NS5- and D5-branes in the corresponding location, hence obey a quantization condition
\beqa
n_a\equiv 32\pi^2 d_a\in\IZ \quad ,\quad m_b\equiv 32 \pi^2 {\tilde d}_b\in\IZ\, .
\label{quant1}
\eeqa
There is also a RR 5-form flux on the corresponding $\IS^2\times \IS^3$'s, which implies some induced D3-brane charge on the corresponding 5-branes, given by the quantities
\beqa
K_a\equiv 32 \pi k_a\in\IZ \quad,\quad L_b\equiv 32\pi l_b\in\IZ\, .
\label{quant2}
\eeqa
The total amount of 5-form flux, which also pierces the 5-sphere in the asymptotic AdS$_5\times\IS^5$, is thus
\beqa
N=\sum_a n_a K_a+\sum_b m_b L_b
\label{the-n}
\eeqa

The fact that one recovers an asymptotic AdS$_5\times\IS^5$ with RR 5-form flux $N$ given by (\ref{the-n}) can be easily checked by performing the asymptotic expantion of (\ref{the-hs}). We have
\beqa 
h_1= 4\Big( r+\frac 2r \sum_b {\tilde d}_bl_b\Big)\sin\varphi \quad ,\quad h_2= -4\Big( r+\frac 2r \sum_a d_a k_a\Big)\cos\varphi
\label{the-asympt-hs}
\eeqa 
Using (\ref{wnn}), we have constant dilaton, and a 10d metric (\ref{ansatz}) given by
\beqa
ds^2= 8\sqrt{2} R \Big( \frac{dr^2}{r^2}+\frac{r^2}{2R^2}ds^2_{AdS_4} +  \big( d\varphi^2+\sin^2\varphi \,ds_{\IS_1^2}^2 + \cos^2\varphi\, ds_{\IS_2^2}^2\big)
\Big)
\label{the-asympt-ads}
\eeqa 
The first two terms describe an AdS$_5$, while those in brackets correspond to an $\IS^5$. Their radius $R$ is related to (\ref{the-n}) by the usual holographic relation $R^4\sim N^2$.

The parameters of the ETW configuration will receive a natural interpretation in terms of the holographic dual, in the next section. We note that, although the parameters $k_a$, $l_b$, $d_a$, ${\tilde b}_b$ are subject to the quantization conditions (\ref{quant1}), (\ref{quant2}), we treat them as continuous in the supergravity approximation if needed.

The interpretation of these geometries is the following. For large $r=|w|$, the geometry asymptotes to AdS$_5\times\IS^5$ with $N$ units of RR 5-form flux.  For smaller values of $|w|$, there is a non-trivial warping, and the solution ends at the origin $w=0$. The solution thus describes an effective end of the world (ETW) configuration for AdS$_5\times\IS^5$, described by a `fat brane' with AdS$_4$ symmetry. Hence it corresponds to a string theory version of the Karch-Randall branes \cite{Karch:2000ct}. 

\begin{figure}[htb]
\begin{center}
\includegraphics[scale=.3]{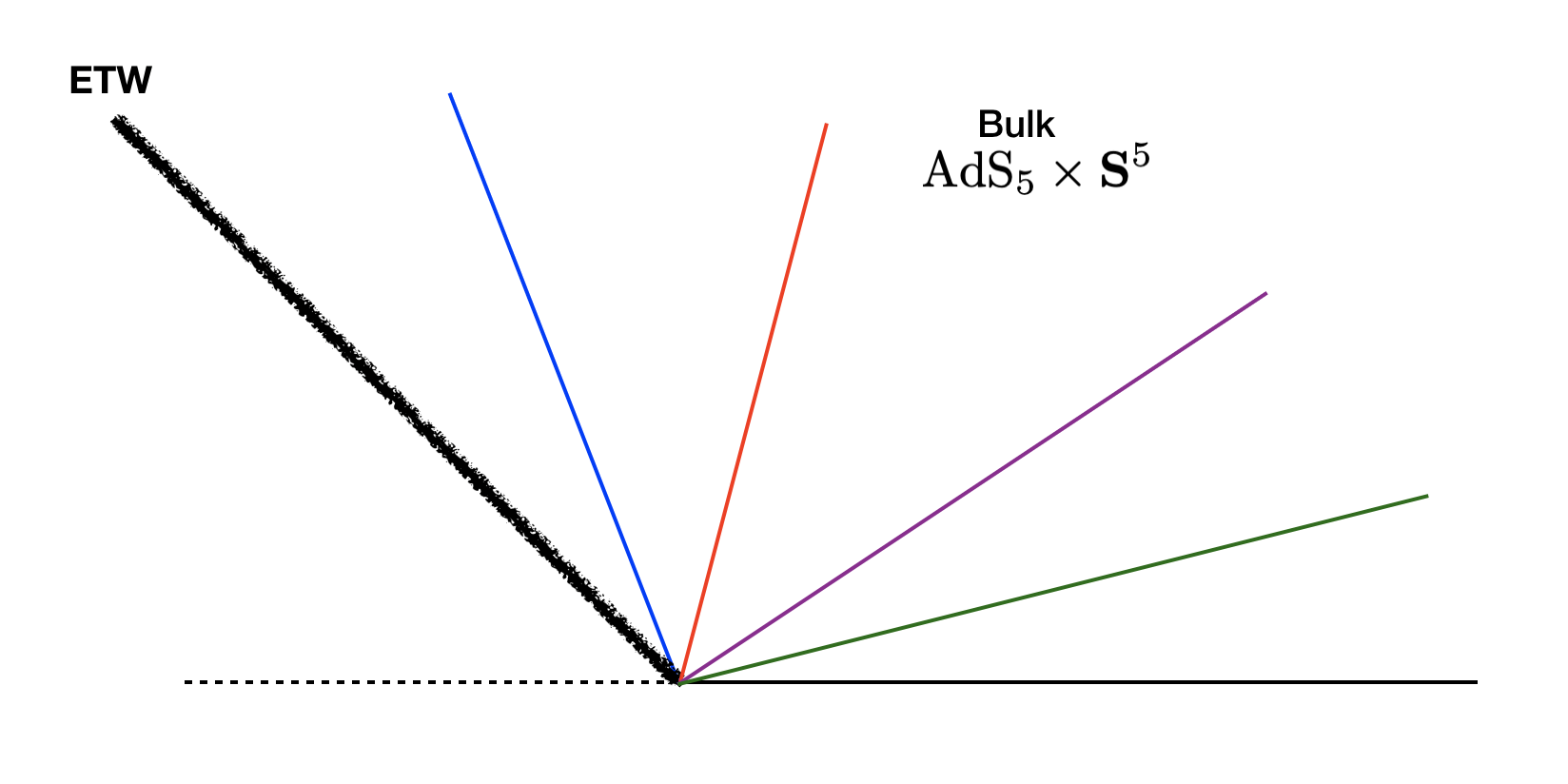}
\caption{\small The geometry shown as a piece of Poincar\'e AdS$_5\times\IS^5$ ending on the ETW brane. The colored radial lines correspond to the arcs of Figure \ref{fig:fivebranes} (inspired in Fig. 4 of \cite{Raamsdonk:2020tin}).}
\label{fig:poincare}
\end{center}
\end{figure}

A key property of these ETW branes is that they extend off to the holographic boundary of the AdS$_5$ geometry, which in fact, engulfs the holographic boundary of the AdS$_4$. This in fact sets the stage for the (double) holographic interpretation in the next section. A picture of the geometry displaying the ETW brane reaching to the holographic boundary is shown in Figure \ref{fig:poincare}

\subsection{Boundaries in 4d $\NN=4$ SYM}
\label{sec:holobound}

In this section we review the holographic description of the solutions in section \ref{sec:case-of-one}.

As already mentioned, the ETW configurations in AdS$_5$ in the previous section reach out to the holographic boundary. Hence, they define the gravity dual of 4d $\NN=4$ $SU(N)$ SYM in a semi-infinite 4d spacetime, namely with a 3d boundary defect. The boundary defect is itself a 3d $\NN=4$ CFT, and the configuration ending the AdS$_5\times \IS^5$ provides its 4d gravity dual, in the picture known as {\em double holography} \cite{Karch:2000gx,Takayanagi:2011zk} (see \cite{Karch:2022rvr} for a recent discussion in this context).
The boundaries preserve half of the super(conformal) symmetry, and are of the kind considered in \cite{Gaiotto:2008ak}, which can be described using Hanany-Witten brane configurations \cite{Hanany:1996ie}. In the following we review the basic ingredients needed for our discussion.

Consider $N$ D3-branes along the directions 0123 and located at the same position in the other directions. We can define a boundary for this theory by letting the D3-branes end on suitable configurations of NS5- and D5-branes localized in the direction 3, and spanning 012, and some additional directions. In order to preserve the maximum supersymmetry, the NS5-branes span the directions 012$\,$456 (and are located in the same position as the D3-branes in 789), while the D5-branes span the directions 012$\,$789 (and are located in the same position as the D3-branes in 456). We consider all 5-branes to be located on the same position in the direction 3, so that they define a boundary for the whole 4d $SU(N)$ theory.\footnote{We note that the 5-branes are taken to the same position in 3 starting from a setup where they maintain a specific ordering, which encodes the specifics of the boundary configuration. This has its reflection in the ordering of 5-brane poles supergravity solution, but we will skip its discussion in the holographic field theory dual context, referring the reader to \cite{Gaiotto:2008ak} for details.}

As studied in \cite{Gaiotto:2008ak}, the boundary conditions of D3-branes ending on a 5-brane (NS5- and D5-branes admit a similar description, due to S-duality) are associated to a representation of $SU(2)$, which describes the commutation relations of the  three (matrix-valued) complex scalars of the 4d SYM, namely $[X^i,X^j]\sim \epsilon^{ijk} X^k$. 
Hence, if we consider a configuration with $N_5$ NS5-branes and $N_5'$ D5-branes, we need to partition each of these quantities in a number of copies of the different irreducible representations of $SU(2)$. In other words
\beqa
{\cal R}_n=\oplus_{a=1}^n n_a{\cal R}_{K_a}\quad , \quad {\cal R}_m=\oplus_{b=1}^m m_b{\cal R}_{L_b}\, ,
\eeqa
where ${\cal R}_{K_a}$, ${\cal R}_{L_b}$ correspond to the irreducible representations of $SU(2)$ of dimensions $K_a$, $L_b$, respectively (here clearly $K_a\neq K_{a'}$ for $a\neq a'$, and similarly for the $L_b$'s).

In other words, we split the NS5-branes into $n$ `stacks', with multiplicity $n_a$, $a=1\ldots, n$, with each NS5-brane in the $a^{th}$ defining boundary conditions associated to the irreducible representation of $SU(2)$ of dimension $K_a$; namely, there are $K_a$ D3-branes ending on each of the $n_a$ NS5-branes in the $a^{th}$ stack. And similarly for the D5-branes. Clearly, to define a boundary for the 4d $\NN=4$ $SU(N)$, all the D3-branes must end on some 5-brane, hence
\beqa
N=\sum_a n_aK_a + \sum m_bL_b\, ,
\eeqa
which matches (\ref{the-n}).
The double holographic interpretation is clear. The boundary configuration is a Hanany-Witten \cite{Hanany:1996ie} set  of D3-branes suspended among NS5- and D5-branes, describing a 3d $\NN=4$ CFT, in the particular limit of all NS5- and D5-branes coincident in the direction 3 (strong coupling limit of the theory at the origin of its (completely) Higgsed phase). Hence, the  AdS$_4$ ETW brane ending the AdS$_5$ ($\times \IS^5$) spacetime is just the gravitational dual of this 3d CFT (see \cite{Assel:2011xz,Assel:2012cj} for this particular class of AdS$_4$/CFT$_3$ duals). The relation of the parameters of the 4d CFT and its 3d boundary CFT determine the Poincar\'e angle formed by the holographic boundary and the ETW boundary, as discussed in detail in \cite{Raamsdonk:2020tin, VanRaamsdonk:2021duo}. Note that, the latter reference considered the regimes where this angle is large; in contrast, we impose no constraint on the parameters and consider general configurations. 

Note that a similar description applies to the gravity solutions with two asymptotic AdS$_5\times \IS^5$ regions (Janus configurations), even with different numbers $N_1$, $N_2$ of units of RR 5-form flux, in section \ref{sec:case-of-one}. Consider without loss of generality that $N_1>N_2$, and denote $N=N_1-N_2$. The holographic description corresponds to $N_2$ infinite D3-branes, intersected by the set of NS5- and D5-branes, and $N$ semi-infinite D3-branes on top; the latter end on the 5-brane configuration with a similar split into $SU(2)$ irreducible representations. 

\section{Dynamical cobordism for AdS$_5\times \IS^5$}
\label{sec:cobordism}

\subsection{The bagpipe geometries are dynamical cobordisms}
\label{sec:bagpipe-cobordism}

The bagpipe geometries in the previous section allow to connect AdS$_5\times \IS^5$ compactifications in a cobordism, which is in fact realized in terms of a fully supersymmetric configuration.

An important point in these statements is that the configurations may in general include regions corresponding to NS5- or D5-branes, as discussed in section \ref{sec:bagpipes}. One may fear that that they lead to additional non-compact spikes which spoil the actual cobordism between AdS asymptotic spaces, basically as in the problematic cobordism with D3-branes mentioned in the introduction. However, it is easy to check that there is no such problem, and that the local geometries including the backreaction of the 5-branes and the induced D3-brane charge are actually compact. Indeed, it was shown in \cite{Aharony:2011yc} that the 10d metric around a 5-brane location (taken as ${\tilde r}=0$) has the form
\beqa 
ds^2=f_4^2\big[ds_{AdS_4}^2+ds_{\IS^2_2}^2\big]+4{\tilde r}^2\rho^2\Big[\,d\psi^2+\sin^2\psi \,ds_{\IS_1^2}^2+\frac{1}{{\tilde r}^2}d{\tilde r}^2\, \Big]\, ,
\eeqa 
with
\beqa 
\rho^2\sim {\tilde r}^{-\frac 32}|\log {\tilde r}|^{-\frac 14}\quad ,\quad f_4^2\sim {\tilde r}^{\frac 12}|\log {\tilde r}|^{\frac 34}\, .
\eeqa
The local geometry is AdS$_4\times \IS^2\times\IS^3$ fibered over the radial direction ${\tilde r}$, which ends at a singularity at ${\tilde r}=0$, at finite distance in spacetime. The singularity is simply due to the presence of an explicit source in the configuration, but it leads to no non-compactness. 

Hence the solutions describe interpolating solutions describing cobordisms among AdS$_5\times\IS^5$ regions.  It is clear that being able to connect arbitrary numbers of AdS$_5\times \IS^5$ pipes actually implies that the cobordism class of a single AdS$_5\times \IS^5$ is trivial. This is in perfect agreement with the swampland cobordism conjecture\footnote{This picture is amusingly reminiscent of the (admittedly non-supersymmetric) solutions in \cite{Hellerman:2010dv}.}.

In particular, this must imply that any two AdS$_5\times \IS^5$ regions can be connected by some domain wall. This indeed corresponds to the configurations with two asymptotic  AdS$_5\times \IS^5$ regions in section \ref{sec:case-of-one}. Similarly, triviality in cobordism must imply that a single AdS$_5\times \IS^5$ admits a boundary ending spacetime; this is again indeed achieved by the geometry with a single asymptotic AdS$_5\times \IS^5$ in section \ref{sec:case-of-one}. 

The configurations with a single AdS$_5\times\IS^5$ spacetime ending on an ETW configuration, when regarded from the 5d perspective, has the structure of a dynamical cobordism, in the sense of \cite{Buratti:2021yia,Buratti:2021fiv,Angius:2022aeq,Blumenhagen:2022mqw,Blumenhagen:2023abk}.
Indeed, the solution corresponding to (\ref{the-hs}) describes, for large $r\equiv |w|\gg \alpha_{g+1},\beta_1$, an asymptotic AdS$_5\times \IS^5$. As $|w|$ decreases towards the region  $r\simeq \alpha_{g+1},\beta_1$, there is a non-trivial running of the solution as we move along the AdS$_4$ slices of the foliation. In particular, considering the $\IS^5$ obtained by fibering the $\IS_1^2\times\IS_2^2$ over the segment $w=re^{i\varphi}$, $\pi/s\leq \varphi\leq \pi$, 
its geometry varies along $r$ (it squashes while preserving the $SO(3)^2$ symmetry of the $\IS^2$'s); similarly, the dilaton also runs with $r$. The discussion of this running solution, from the perspective of the 5d theory, is described in section \ref{sec:eft}. Finally, for $r\ll l_1^2,k_n^2$, the $\IS^5$ becomes essentially round again, and shrinks to zero size, closing off the geometry and ending spacetime. In section \ref{sec:scalars} we discuss this small $r$ behaviour in term of the local dynamical cobordism of \cite{Angius:2022aeq}.

In the next section we expand of this interpretation of the solutions as dynamical cobordisms of the 5d theory resulting after reduction of type IIB theory on the $\IS^5$ coordinates. Note that the interpretation of these solutions from the lower-dimensional perspective aligns with the viewpoint implicit in holography. Hence, we expect our approach to relate to the bottom up descriptions in the study of AdS/BCFT and quantum islands. 

\subsection{A 5d effective theory description}
\label{sec:eft}

The actual description of the above solutions as dynamical cobordisms requires describing them as 5d running solutions ending in a 4d ETW configuration. In this section we develop such 5d effective description. 

We recall that, since already in AdS$_5\times\IS^5$ there is no scale separation (in agreement with the AdS distance conjecture \cite{Lust:2019zwm}), this 5d theory should not be regarded as an effective field theory in the Wilsonian sense, but rather as an $\IS^5$ truncation. Namely, we perform a reduction along the $\IS^5$ keeping a finite set of modes which capture the essential dynamics of the full configuration. As is well known, for AdS$_5\times\IS^5$ this corresponds to a consistent truncation, and the dynamics of the $SO(6)$ invariant sector has little mixing with the KK modes, despite the lack of mass hierarchy in the KK tower. For the configuration with ETW branes, the lower symmetry implies a more significant role of the KK modes involved in the breaking $SO(6)\to SO(3)\times SO(3)$. We will first describe the physics of the $SO(6)$ invariant truncation, and show that it is the key player in the ending of spacetime; we subsequently discuss the symmetry breaking modes, and show that they manifest as ETW brane worldvolume modes. For a recent discussion of scale separation in this and other contexts of gravity localization, see \cite{DeLuca:2023kjj}.

\subsubsection{The $\IS^5$ reduction of the metric}
\label{sec:reduct}

We start the discussion of the $SO(6)$ invariant sector, and show it suffices to capture the ending of the AdS$_5$ spacetime on a smoothed out version of an AdS$_4$ Karch-Randall brane.
The ETW configuration is to be described as a running solution of a 5d theory including the 5d metric, the IIB complex dilaton (although we restrict the discussion to vanishing axion), and a breathing mode of a round $\IS^5$. As explained, in the next section we enrich the system by the inclusion of higher KK modes, describing the squashing $SO(6)\to SO(3)^2$ of the $\IS^5$.

Consider the 10d ETW solutions in section \ref{sec:case-of-one}, whose the 10d metric (\ref{ansatz}) we repeat for convenience
\beqa
ds^2=f_4^2 ds_{AdS_4}^2+f_1^2ds_{\IS^2_1}^2+f_2^2ds_{\IS_2^2}^2+4\rho^2(dr^2+r^2d\varphi^2)
\eeqa
where have have used polar coordinates for the Riemann surface $\Sigma$. Recall that the $\IS^5$ is obtained by fibering the $\IS_1^2\times \IS_2^2$ over the interval $\pi/2\leq\varphi\leq \pi$ at fixed value of the radial coordiante $r$

Hence, we carry out this reduction in two steps. First, we restrict to the $SO(3)\times SO(3)$ invariant sector by performing the reduction on $\IS^2_1\times\IS^2_2$, by using the general ansatz
\beqa
ds^2=e^{-\frac{1}{2}(\sigma_1+\sigma_2)}ds_6^2+e^{\frac{1}{\sqrt{2}}(a\sigma_1-b\sigma_2)}ds_{\IS^2_1}^2+e^{\frac{1}{\sqrt{2}}(a \sigma_2-b\sigma_1)}ds_{\IS_2^2}^2\, ,
\eeqa
where $a,b$ will be fixed shortly. The resulting 6d metric is
\beqa
ds_6^2=\sqrt{f_1^2f_2^2}\, [\,f_4^2 ds_{AdS_4}^2+4\rho^2(dr^2+r^2d\varphi^2)\,]\, .
\label{metric-6d}
\eeqa
We also have two 6d scalars $\sigma_1$, $\sigma_2$ with profiles
\beqa
\sigma_1=\log(f_1^af_2^b) \quad,\quad \sigma_2=\log(f_1^bf_2^a) \;,\; {\rm with}\, a=1+\frac 1{\sqrt{2}}\; ,\; b= 1-\frac 1{\sqrt{2}}\, ,
\eeqa
where the parameters $a,b$ are fixed so that these scalars have canonical kinetic term in the resulting 6d action, although we will not need it. 

Using  (\ref{wnn}), (\ref{dilaton}), (\ref{the-fs}), the metric components can be recast as
\beqa 
\sqrt{f_1^2f_2^2}\,f_4^2=4h_1h_2\quad , \quad \sqrt{f_1^2f_2^2}\,\rho^2=2|W|
\label{6d_ads}
\eeqa 
Using the harmonic function (\ref{the-hs}), they are
\begin{equation}\label{6d_ads_ds}
    \begin{split}   
    \sqrt{f_1^2f_2^2}\,f_4^2= & -32r^2\sin2\varphi-32r\cos\varphi\sum^{m}_{b=1} \tilde{d}_b \log \left( \frac{r^2+l_b^2+2rl_b\sin\varphi}{r^2+l_b^2-2rl_b\sin\varphi} \right)\\
    & -32r\sin\varphi\sum^{n}_{a=1} d_a \log \left( \frac{r^2+k_a^2+2rk_a\cos\varphi}{r^2+k_a^2-2rk_a\cos\varphi} \right)\\ 
    & -16\sum^{n}_{a=1}\sum^{m}_{b=1} d_a\tilde{d}_b \log \left( \frac{r^2+k_a^2+2rk_a\cos\varphi}{r^2+k_a^2-2rk_a\cos\varphi} \right)\log \left( \frac{r^2+l_b^2+2rl_b\sin\varphi}{r^2+l_b^2-2rl_b\sin\varphi} \right)
    \end{split}
\end{equation}
\begin{equation}
\begin{split}
    \sqrt{f_1^2f_2^2}\,\rho^2= & -32r^2\sin2\varphi \left[\sum^{n}_{a=1}\frac{d_ak_a}{(r^4+k_a^4-2r^2k_a^2\cos 2\varphi)}+
    \sum^{m}_{b=1}\frac{\tilde{d}_bl_b}{(r^4+l_b^4+2r^2l_b^2\cos 2\varphi)}+\right.\\
    & +2 \left. \sum^{n}_{a=1}\sum^{m}_{b=1}\frac{d_a\tilde{d}_bk_al_b(k_a^2+l_b^2)}{(r^4+k_a^4-2r^2k_a^2\cos 2\varphi)(r^4+l_b^4+2r^2l_b^2\cos 2\varphi)}\right]
\end{split}
\label{6d_radius_ds}
\end{equation}

For large $r$ the 6d metric behaves as 
\beqa
ds^2=-64R^2 \sin 2 \varphi \bigg(\, \frac{dr^2}{r^2}+\frac{r^2}{2R^2}ds_{AdS_4}^2+d\varphi^2\bigg)\, .
\label{ads-6d-general}
\eeqa
This clearly includes an AdS$_5$ factor, and an $\IS^1$ left over from $\IS^5$ upon integration over $\IS^2\times\IS^2$, of radius
\beqa 
R^2=\sum^{n}_{a=1}(\beta_a-k_a^2)-\sum^{m}_{b=1}(\alpha_{b+n+1}+l_b^2)
\eeqa 
We recall that this relates to the number of D3 branes $N$ by the usual AdS/CFT dictionary.

The interpretation of the $\varphi$-dependent prefactor in (\ref{ads-6d-general}) is simply the zero mode $f_0$ in the KK tower of the 5d graviton. Fixing the normalization $\int_{\pi/2}^\pi f_0^2=1$, we have
\begin{equation}
f_0(\varphi)=-\frac{2}{\sqrt{\pi}}\sin 2\varphi\, ,
\end{equation}
where the minus sign is introduced for later convenience.
This allows to define the $\IS^1$ truncation onto the $SO(6)$ invariant sector. Since KK modes form an orthonormal basis, the 5d avatar of the full 6d metric can be extracted as
\beqa 
\hat{g}_{\mu\nu}=\int_{\pi/2}^\pi f_0(\varphi) g_{\mu\nu}(r,\varphi)d\varphi
\eeqa
This gives 
\beqa
ds_5^2=\hat{f}_4^2 ds_{AdS_4}^2+4\hat{\rho}^2\, dr^2\, .
\label{metric-5d}
\eeqa
with
\begin{equation}\label{0mode_ads4}
    \begin{split}   
    \hat{f}_4^2= & 16\sqrt{\pi}r^2+32\sqrt{\pi}r\sum^{m}_{b=1} \tilde{d}_b\left(B_b-\frac{B_b^3}{3}\right) +32\sqrt{\pi}r\sum^{n}_{a=1} d_a\left(A_a-\frac{A_a^3}{3}\right)+ \\ 
    & +64\sqrt{\pi}\sum^{n}_{a=1}\sum^{m}_{b=1} d_a\tilde{d}_b \left[A_aB_b+\frac{A_a^2+B_b^2}{2A_a^2B_b^2}\left(A_aB_b-(1+A_a^2B_b^2)\tan^{-1}(A_aB_b)\right)\right]
    \end{split}
\end{equation}

\begin{equation}\label{0mode_rho}
    \hat{\rho}^2= 16\sqrt{\pi} \left[\sum^{n}_{a=1}\frac{d_a}{k_a}A_a^2+\sum^{m}_{b=1}\frac{\tilde{d}_b}{l_b}B_b^2+2 \sum^{n}_{a=1}\sum^{m}_{b=1}d_a\tilde{d}_bk_al_b\frac{A_a^2+B_b^2}{r^4+k_a^2l_b^2}\right]
\end{equation}
where 
\beqa 
&&A_a=\left\{\begin{array}{cc}
 r/k_a, &  \ \  r\leq k_a \\ 
 k_a/r, &  \ \  r>k_a \\
\end{array}\right. \quad , \quad
B_b=\left\{\begin{array}{cc}
 r/l_b, &  \ \  r\leq l_b \\ 
 l_b/r, &  \ \  r>l_b 
\end{array}\right.\, .
\eeqa 

In order to compare with KR brane, it is convenient to change coordinates to gather all the non-trivial structure of the metric in a unique warp factor $A$ in front of the AdS$_4$ piece. We choose the parametrization
\beqa\label{5d_metric_KR}
ds_{5d}^2=e^{2A(x)} ds_{AdS_4}^2+\frac{128R^2}{\sum^{n}_{a=1}k_a^2+\sum^{m}_{b=1}l_b^2} dx^2\, ,
\eeqa

Some illustrative plots of the quantities $\hat{f}_4^2$ and $A$, and their implication in the KR localization of gravity are discussed in section \ref{sec:localized}.

\subsubsection{The single 5-brane example}
\label{sec:single5}

In order to get some insight we now consider the particular case of a single stack of NS5-branes are the position $k$, the resulting factors for the 6d metric (\ref{metric-6d}) are

\beqa
\sqrt{f_1^2f_2^2}\,f_4^2=-32r^2\sin2\varphi - 16\, r\sin\varphi \,  \frac{\beta-k^2}{k} \log \left( \frac{r^2+k^2+2rk\cos\varphi}{r^2+k^2-2rk\cos\varphi} \right)
\eeqa

\beqa
\sqrt{f_1^2f_2^2}\, \rho^2= \frac{-16\, r^2(\beta-k^2)\sin2\varphi}{r^4+k^4-2r^2k^2\cos 2\varphi}
\eeqa

In this case the radius is given by
\beqa 
R^2=\beta-k^2
\eeqa 
For the components of the 5d metric (\ref{metric-5d}) we have

\begin{equation}
    \hat{f}_4^2= 16\sqrt{\pi}\, r^2 +16\sqrt{\pi}\, r\, \frac{\beta-k^2}{k}\left(A-\frac{A^3}{3}\right)
\end{equation}

\begin{equation}
    \hat{\rho}^2=8\sqrt{\pi}\, \frac{\beta-k^2}{k^2}A^2
\end{equation}
with $A=\left\{\begin{array}{cc}
 r/k, &  \ \  r\leq k \\ 
 k/r, &  \ \  r>k \\
\end{array}\right.$.

The metric is continuous at $r = k$, and so are its first derivatives. The curvature is finite but displays a discontinuity
\beqa
\Delta R= R_{r=k^+}-R_{r=k^-}=\frac{-\beta}{2\sqrt{\pi}\, k^2(\beta-k^2)}
\eeqa 

It is easy to check that there is no localized stress-energy tensor at the discontinuity. The jump in the curvature is associated to a jump in the matter stress-energy tensor, ultimately related to the change in the fluxes across the 5-brane location.

It is interesting to express the 5d metric in the parametrization of \cite{Karch:2000ct}. Hence, we recast \eqref{metric-5d} as:
\beqa\label{5d_metric_KR_NS5}
ds_{5d}^2=e^{2A(x)} ds_{AdS_4}^2+128\frac{\beta-k^2}{k^2} dx^2\, ,
\eeqa
where the prefactor of the $dx^2$ term is introduced to simplify the expression for $A(x)$.
Here $x$ relates to the previous $r$ as:
\begin{equation}
    \frac{r^2}{k^2}= \left\{ \begin{array}{cc}
 \frac{4x}{\pi^{1/4}k}, &\, r\leq k \\
 \exp\left(\frac{4x}{\pi^{1/4}k}-1\right), &\, r>k \\
\end{array}\right.
\end{equation}

and there is an analytical, yet admittedly not simple, expression for $A(x)$:
\begin{equation}
    e^{2A(x)}= 16\sqrt{\pi}k^2\left\{ \begin{array}{cc}
 \frac{4x}{\pi^{1/4}k}\left(\frac{\beta}{k^2}-\frac{\beta-k^2}{3k^2}\frac{4x}{\pi^{1/4}k}\right) &\, , \; x\leq \frac{\pi^{1/4}k}{4} \\
 \exp\left(\frac{4x}{\pi^{1/4}k}-1\right)-\frac{\beta-k^2}{3k^2}\exp\left(1-\frac{4x}{\pi^{1/4}k}\right) + \frac{\beta-k^2}{k^2} & \, , \;x>\frac{\pi^{1/4}k}{4} \\
\end{array}\right.
\label{the-a}
\end{equation}

\subsubsection{Localization of gravity}
\label{sec:localized}

In order to get some intuition about the solutions, we provide plots for the warp factors $\hat{f}_4^2$, $\hat{\rho}^2$ and $A(x)$, in a few illustrative examples. The configurations display localized gravity when the shape contains a bump, since it corresponds to a smoothed version of the cusp in the KR branes in \cite{Karch:2000ct}. The smoothing is because our 5d theory is not one of pure gravity, but contains scalar degrees of freedom, to be discussed in the next section.
We also note that the 5d metric becomes singular at $r=0$, reflecting the fact that the $\IS^5$ shrinks at that point, as we discuss in more detail in section \ref{sec:scalars}.

\begin{figure}[htb]
\begin{center}
\includegraphics[scale=.53]{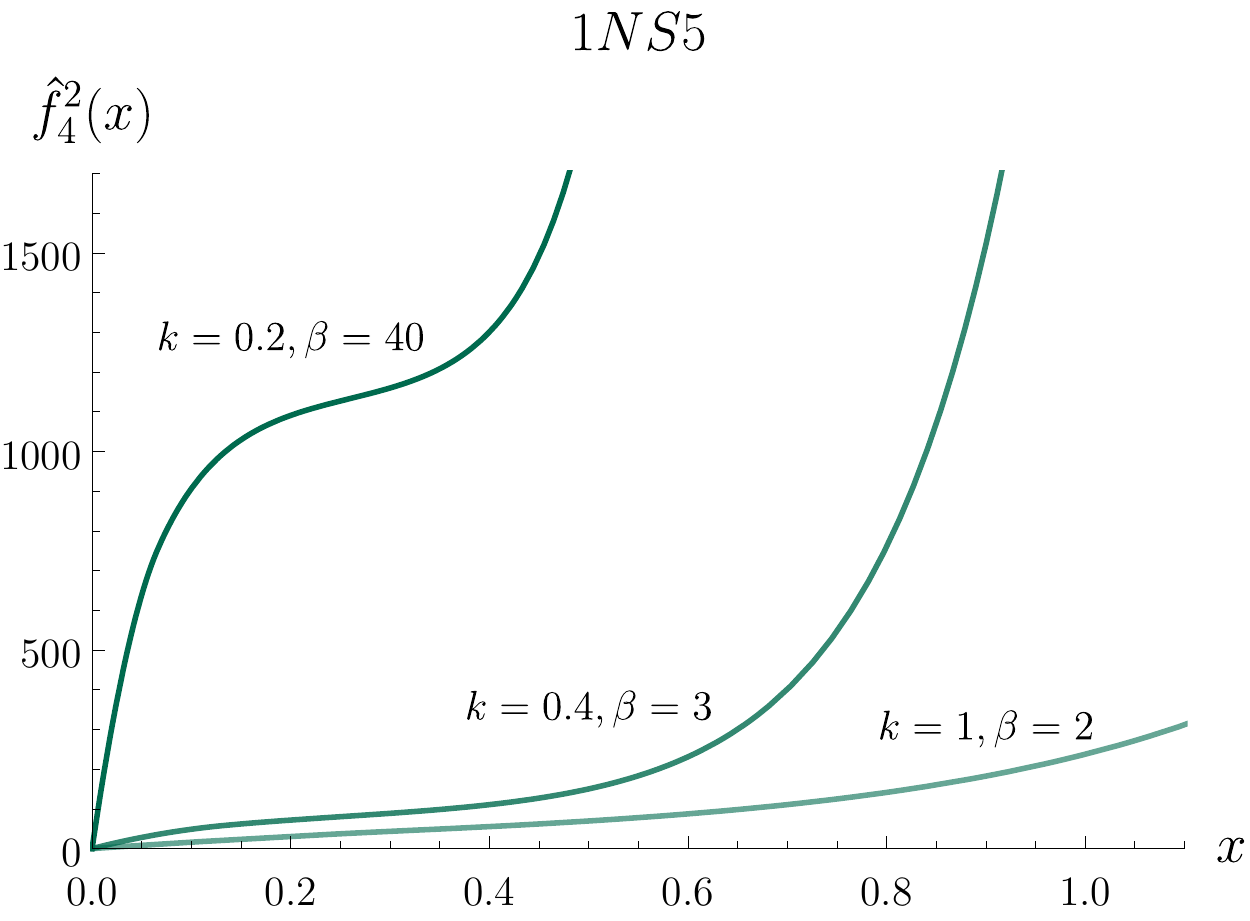}\hspace*{1cm}
\includegraphics[scale=.53]{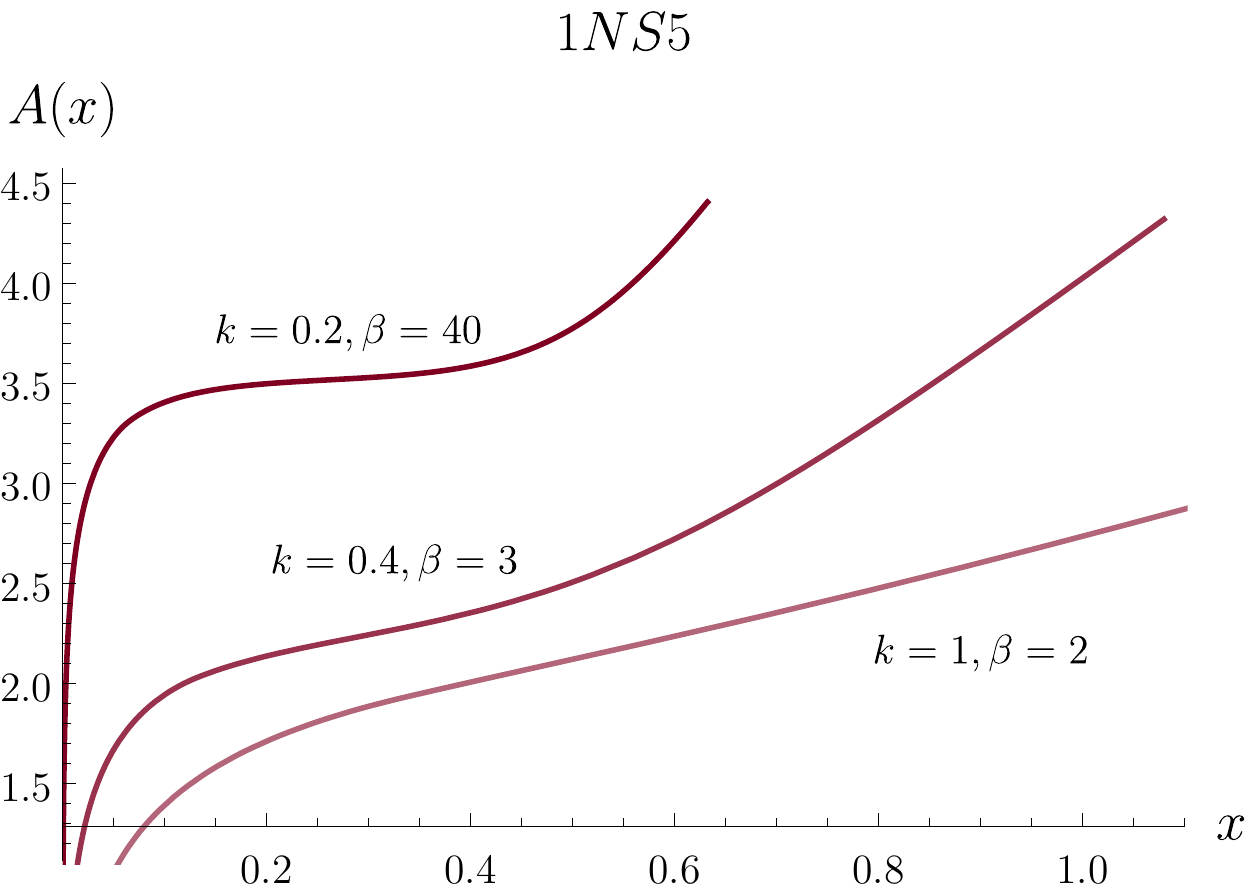}
\caption{\small The warp factors $\hat{f}_4^2$ and $A$ for a single NS5-brane for different values of its parameters. The curve is monotonic and exhibits no bump, hence no localization of gravity.
}
\label{fig:nobump}
\end{center}
\end{figure}

We start considering one single 5-brane. Due to S-duality, we can focus without loss of generality on the case of a single NS5-brane, located at $r=k$. Figure \ref{fig:nobump} shows two illustrative examples of the KR warp factor $A$ in (\ref{the-a}) as a function of the spacetime distance $\Delta$ (proportional to $x$).

The main features are that, as mentioned, the metric and first derivative are continuous across $r=k$. The solution asymptotes to AdS$_5$ at $r\to \infty$ and develops a singularity ($A\to -\infty$) at the origin. But the most specific feature of this case is the fact that it is monotonous and contains no bump. Hence it does not lead to localization of gravity. It is easy to check that this hold for different choices of the parameters, in particular of the relative values of $\beta$ and $k^2$, as illustrated in Figure \ref{fig:nobump}. For $\beta\gg k^2$, the curve flattens out, but develop no bump.

\begin{figure}[htb]
\begin{center}
\includegraphics[scale=.53]{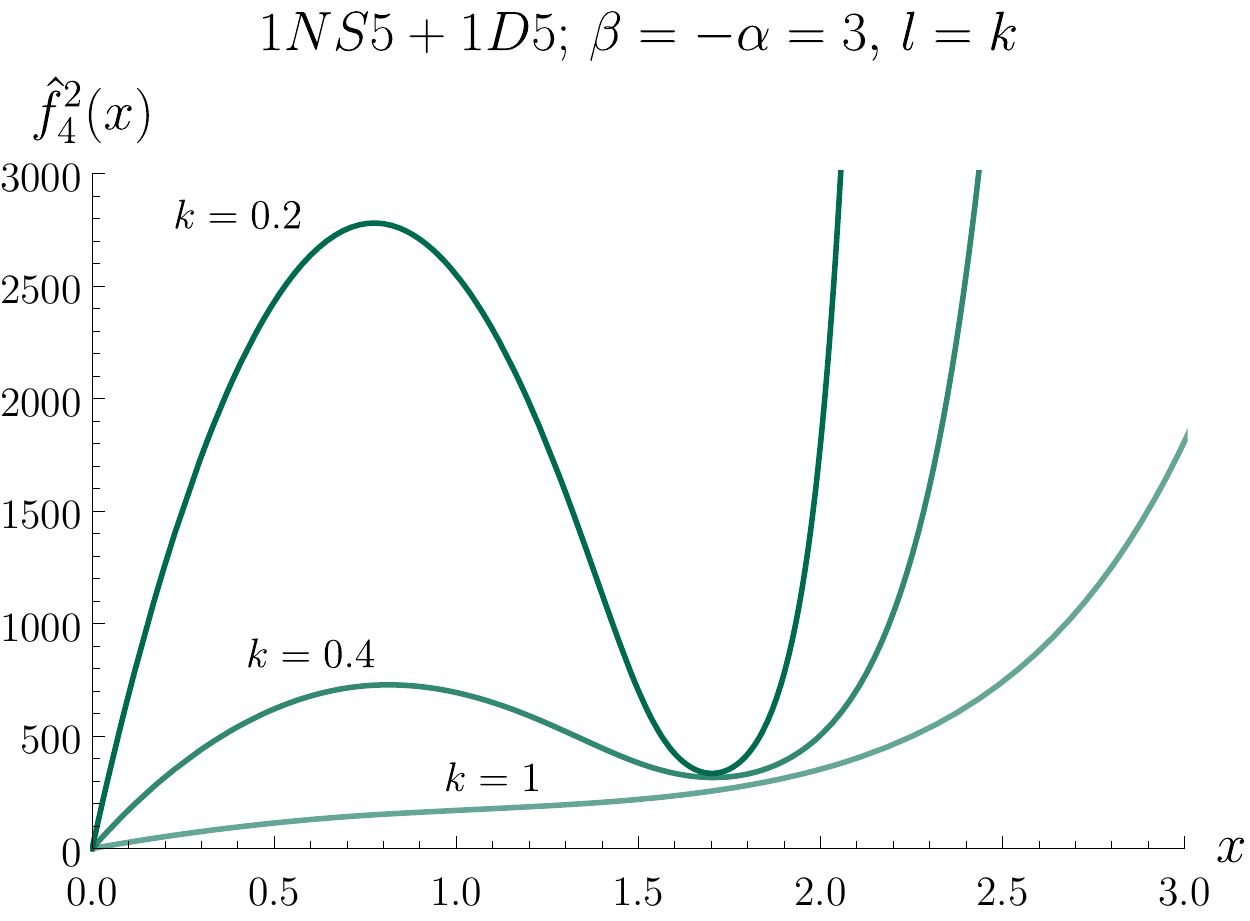}\hspace*{1cm}
\includegraphics[scale=.53]{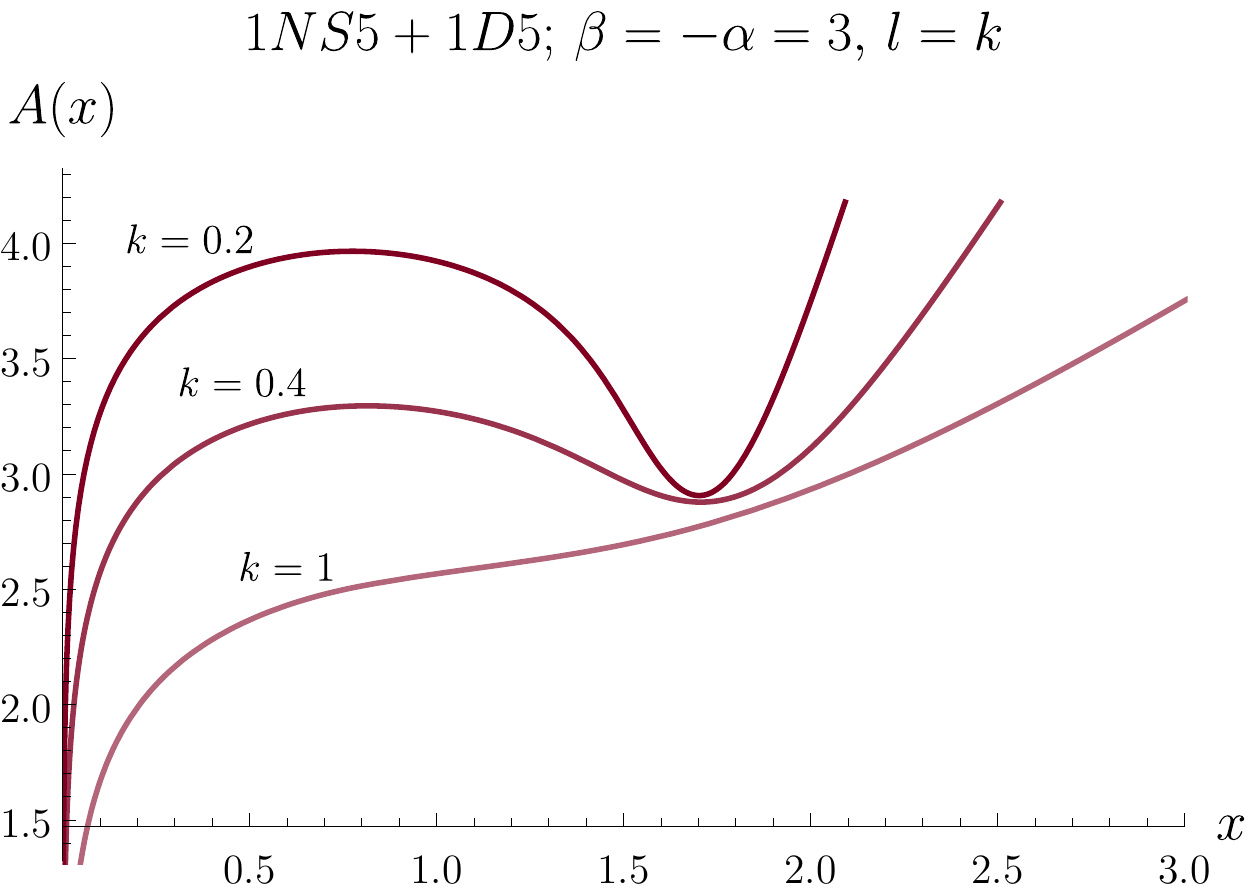}
\caption{\small Warp factors $\hat{f}_4^2$ and $A$ in examples with one NS5- and one D5-branes, with equal parameters, for fixed $\beta$ and different choices of $k$. There is a gravity localizing bump, which gets stronger as $k^2$ becomes smaller.}
\label{fig:change-k}
\end{center}
\end{figure}

It is easy to show that in order to have a bump in the warp factor $A$ and hence localization of gravity, one needs to introduce both kinds of 5-branes simultaneously. We illustrate this in Figure \ref{fig:change-k} in examples with one NS5- and one D5-brane (taking $\alpha=-\beta$, $k=l$, similar results hold also for unequal values, see later). The curves have the same asymptotic features as discussed above, but generically have a bump, leading to a smooth version of the cusp in the bottom-up description of KR branes.

Let us note the interesting pattern upon changing the relative size of $\beta$ and $k^2$, see Figure \ref{fig:change-k}. For small value of $\beta-k^2$ there is no bump even if we have both kinds of 5-branes in the system. On the other hand, as the value of $\beta-k^2$ gets larger, the bump gets stronger, and more decoupled from the asymptotic AdS$_5$. This will be easily understood in Section \ref{sec:double-scaling}.

\begin{figure}[htb]
\begin{center}
\includegraphics[scale=.53]{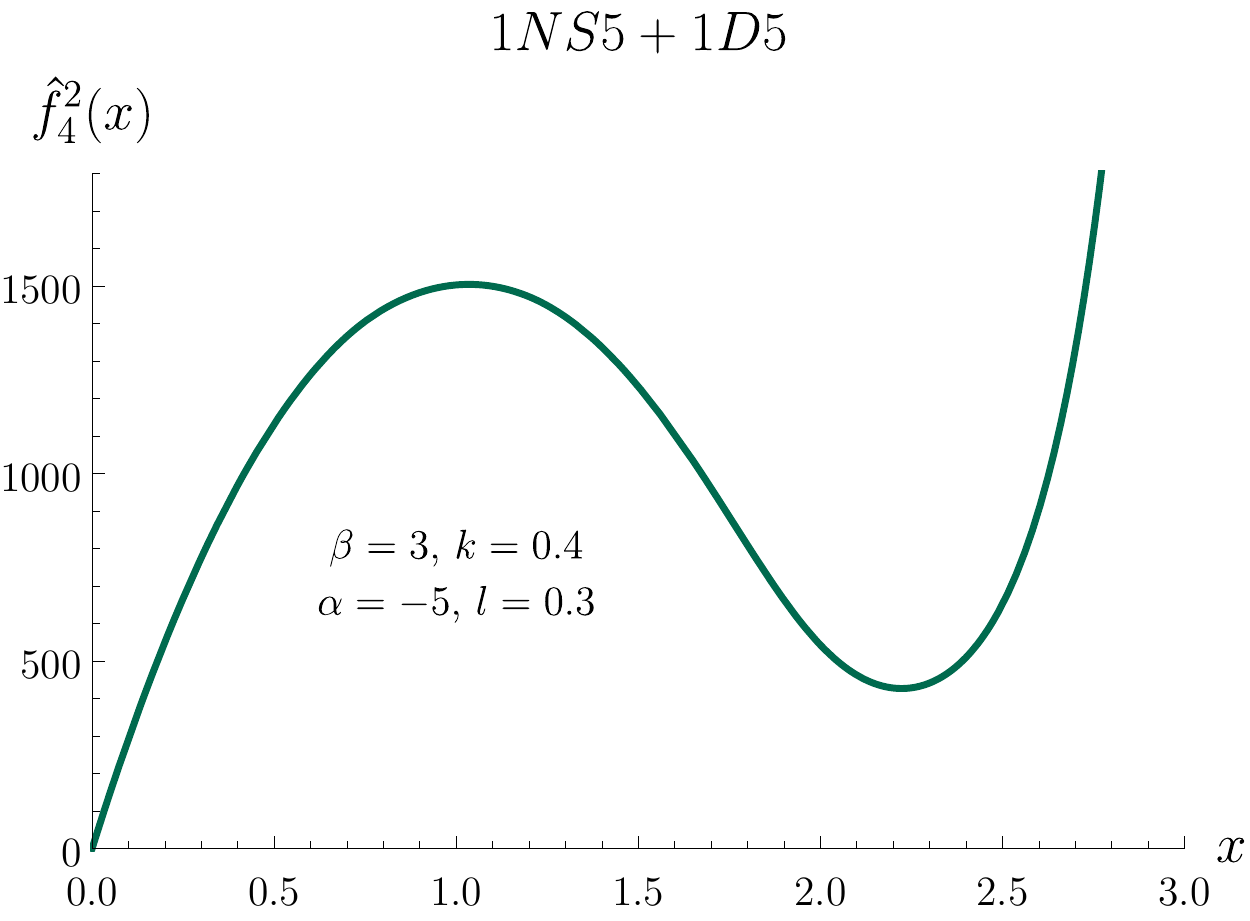} \hspace*{1cm}
\includegraphics[scale=.53]{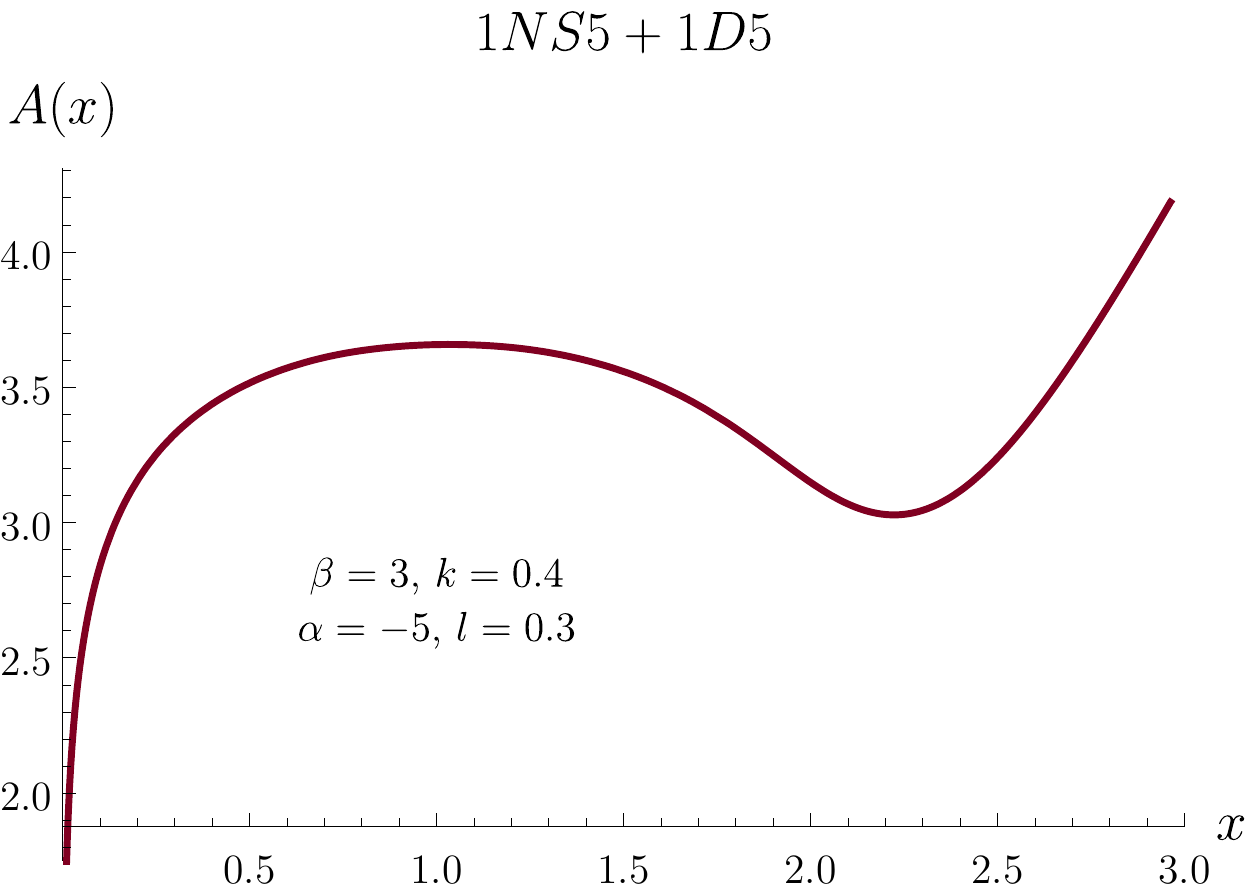}
\caption{\small The bump in the warp factors $\hat{f}_4^4$ and $A$  in an illustrative example with two 5-branes with different parameters.}
\label{fig:different}
\end{center}
\end{figure}

The above results continue to hold even if we allow the 5-branes to have different parameters, see Figure \ref{fig:different} for an illustrative example.

In conclusion, our 5d reduction of the metric reproduces a smooth version of the warp factor of KR branes, in particular with a bump localizing gravity. The smoothing of the cusp of KR branes is due to the presence of additional scalar fields in the 5d theory. They are also responsible for the singularity at the origin, associated to the shrinking of the $\IS^5$, and hence encoded in the dynamics of the its volume modulus, to which we turn next.

\subsubsection{The 5d scalars}
\label{sec:scalars}

\subsubsection*{The $SO(6)$ invariant sector}

There are several scalars that to be included in the 5d description. In this section we focus on the $SO(6)$ invariant sector, namely the dilaton and the $\IS^5$ breathing mode.

The 5d profile of the dilaton is obtained by the reduction of its 10d solution on  $\IS_1^2\times\IS_2^2$, which is trivial, followed by a truncation to the zero mode in the reduction on the $\IS^1$ parametrized by $\varphi$. The zero mode is read out from the asymptotic AdS$_5\times \IS^5$ region, and is simply a constant function. Hence the last $\IS^1$ reduction is simply an average in $\varphi$ of the dilaton profile. 

\begin{figure}[htb]
\begin{center}
\includegraphics[scale=.6]{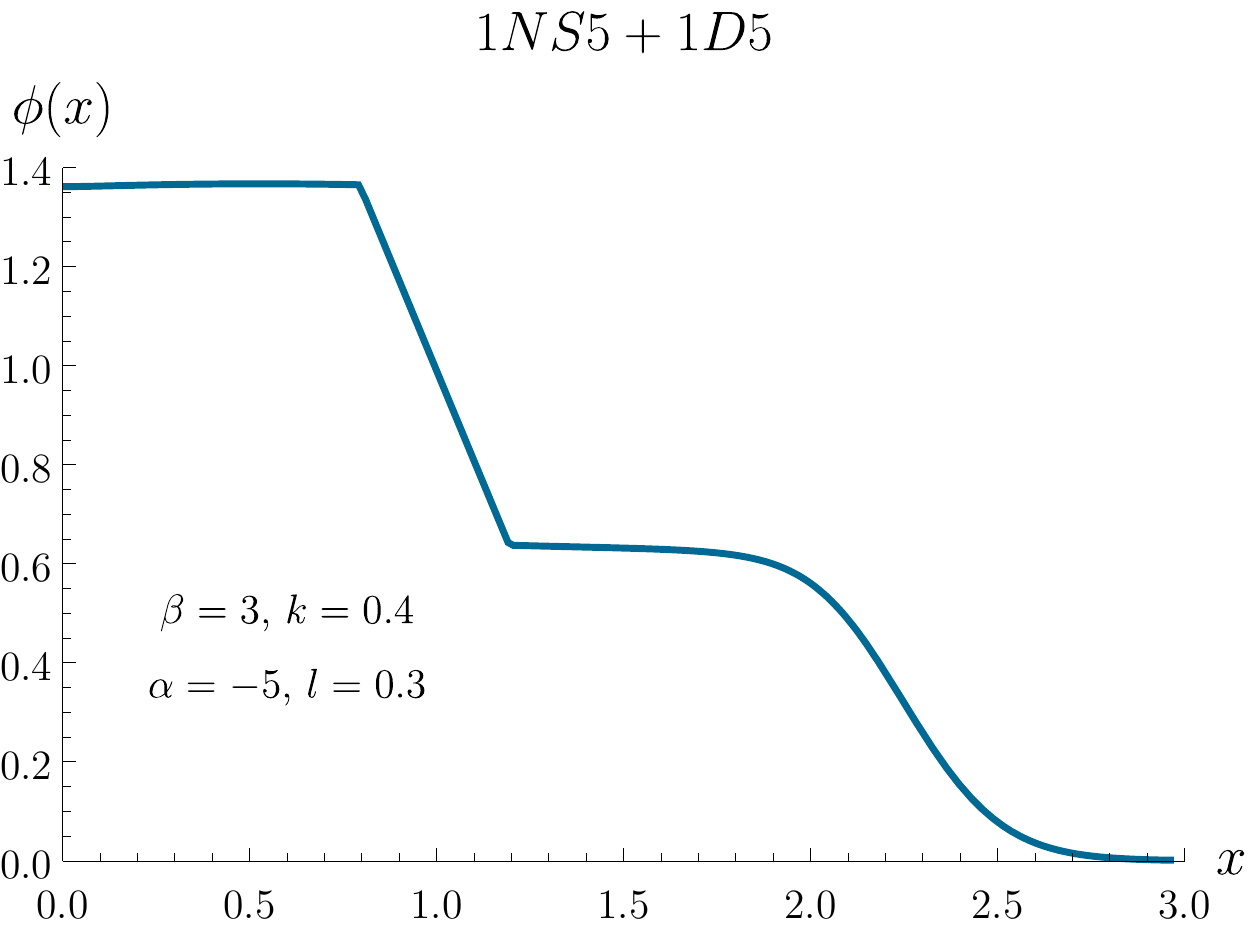}
\caption{\small The dilaton for a  configuration of one NS5- and one D5-brane with different parameters, chose as in Figure \ref{fig:different}. The dilaton behaves as approximately piecewise linear function in the ETW brane region.}
\label{fig:dilaton-one}
\end{center}
\end{figure}

\begin{figure}[htb]
\begin{center}
\includegraphics[scale=.53]{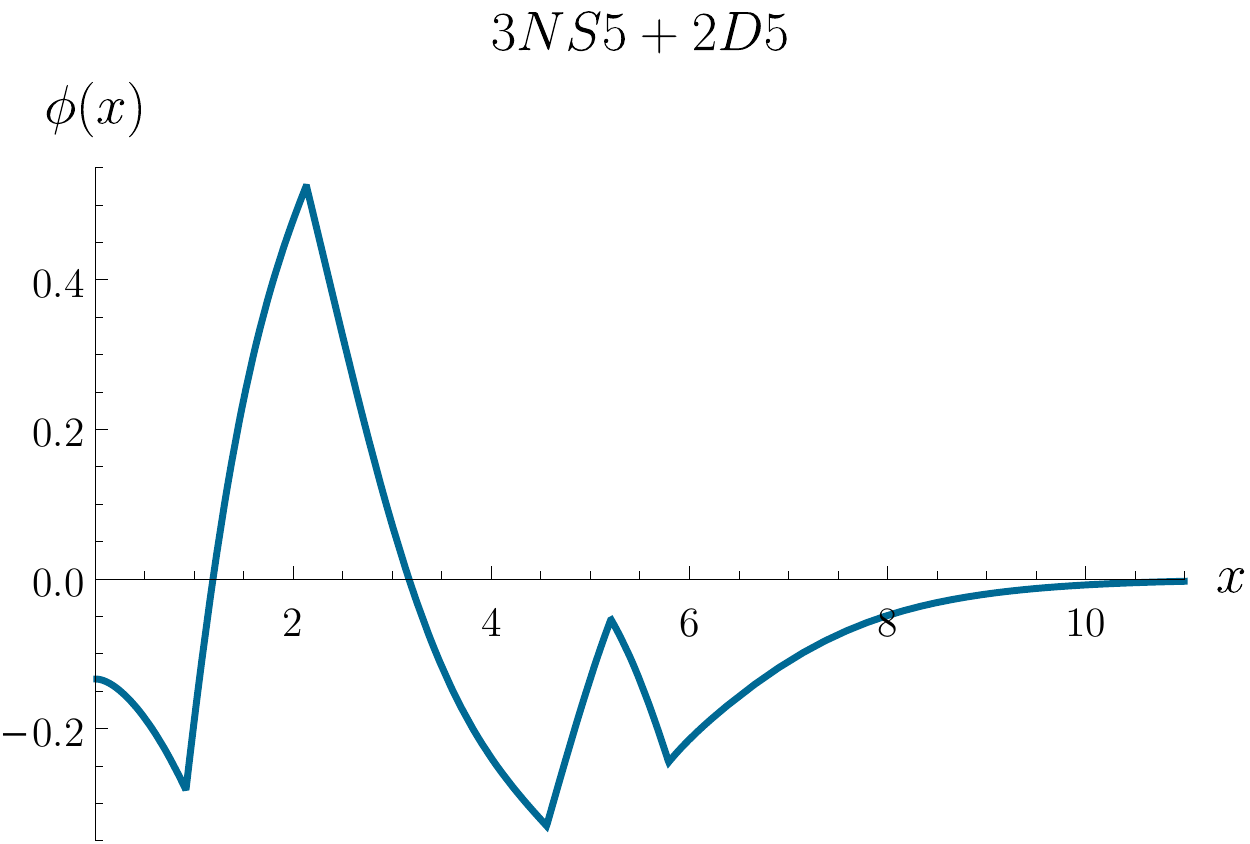}\hspace*{1cm}
\includegraphics[scale=.53]{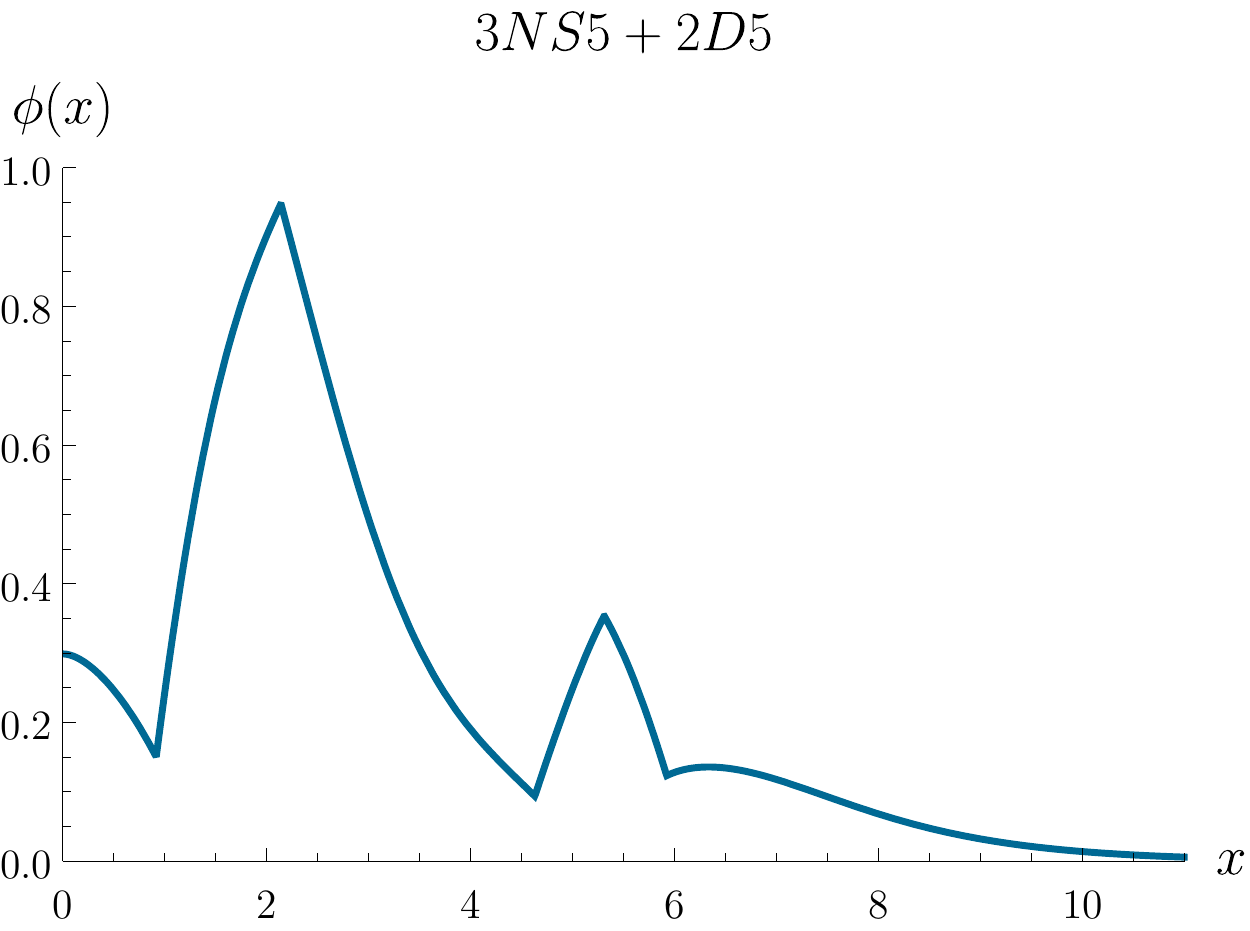}
\caption{\small The dilaton for a fairly general configuration of  NS5- and D5-branes, for parameters $k_1=4$, $k_2=3$, $k_3=0.8$, $\beta_1=22$, $\beta_2=12$, $\beta_3=1.1$, $l_1=1.4$, $l_2=3.5$, $\alpha_5=-4$  (and finally $\alpha_6=-17$ in the first figure, and $\alpha_6=-24$ in the second).}
\label{fig:dilaton-many}
\end{center}
\end{figure}

One interesting observation is that, due to S-duality, the effects of  NS5- and D5-branes on the dilaton have compensating effects.  
Hence, in order to better appreciate the generic behaviour of the dilaton it is useful to consider configurations in which both kinds of 5-branes have different parameters. 
In Figure \ref{fig:dilaton-one} we show an illustrative example with one NS5- and one D5-brane. The dilaton is asymptotically constant and gets perturbed in the vicinity of the 5-branes, which bend its slope in opposite ways for NS- and D5-branes. The dilaton behaves as approximately piecewise linear function. This is however not an exact behaviour, as can be seen in more generic situations, see for instance Figure \ref{fig:dilaton-many}.

\medskip

We now consider the $\IS^5$ breathing mode. In the asymptotic AdS$_5\times\IS^5$ region, this can be characterized by using the 6d scalars  $\sigma_1$, $\sigma_2$ describing the breathing modes of the $\IS^2$'s. In particular one may use any function of them,  invariant under the $1\leftrightarrow 2$ exchange. Different choices can differ in detailed behaviour in the mid region, but this ambiguity is a mere reflection of the non-trivial mixing of KK modes due to the breaking of symmetries at the core of the ETW brane. In the following we proceed with a particular simple choice, warning the reader that the resulting picture is necessarily only reliable at the qualitative level.

\begin{figure}[htb]
\begin{center}
\includegraphics[scale=.53]{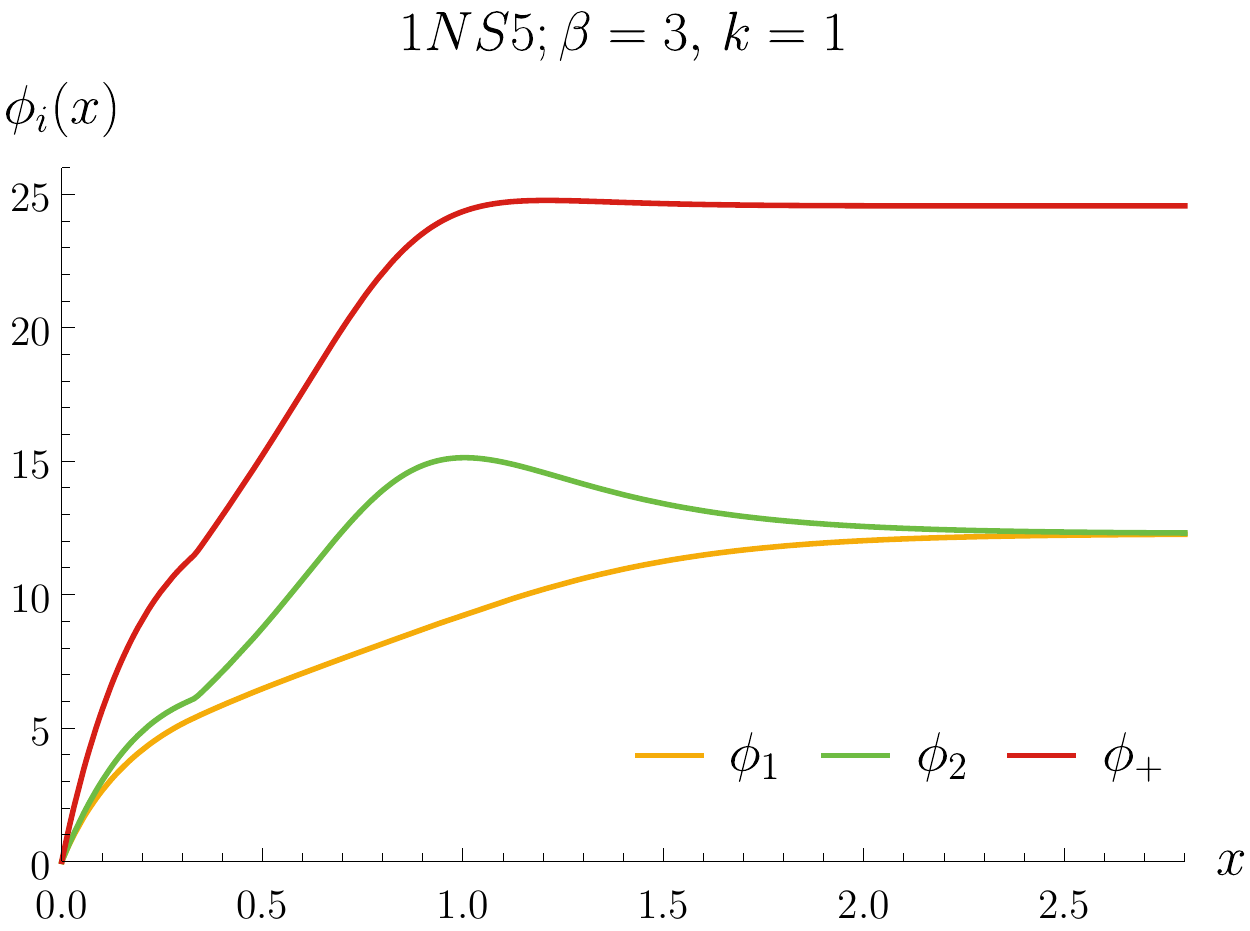}\hspace*{1cm}
\includegraphics[scale=.53]{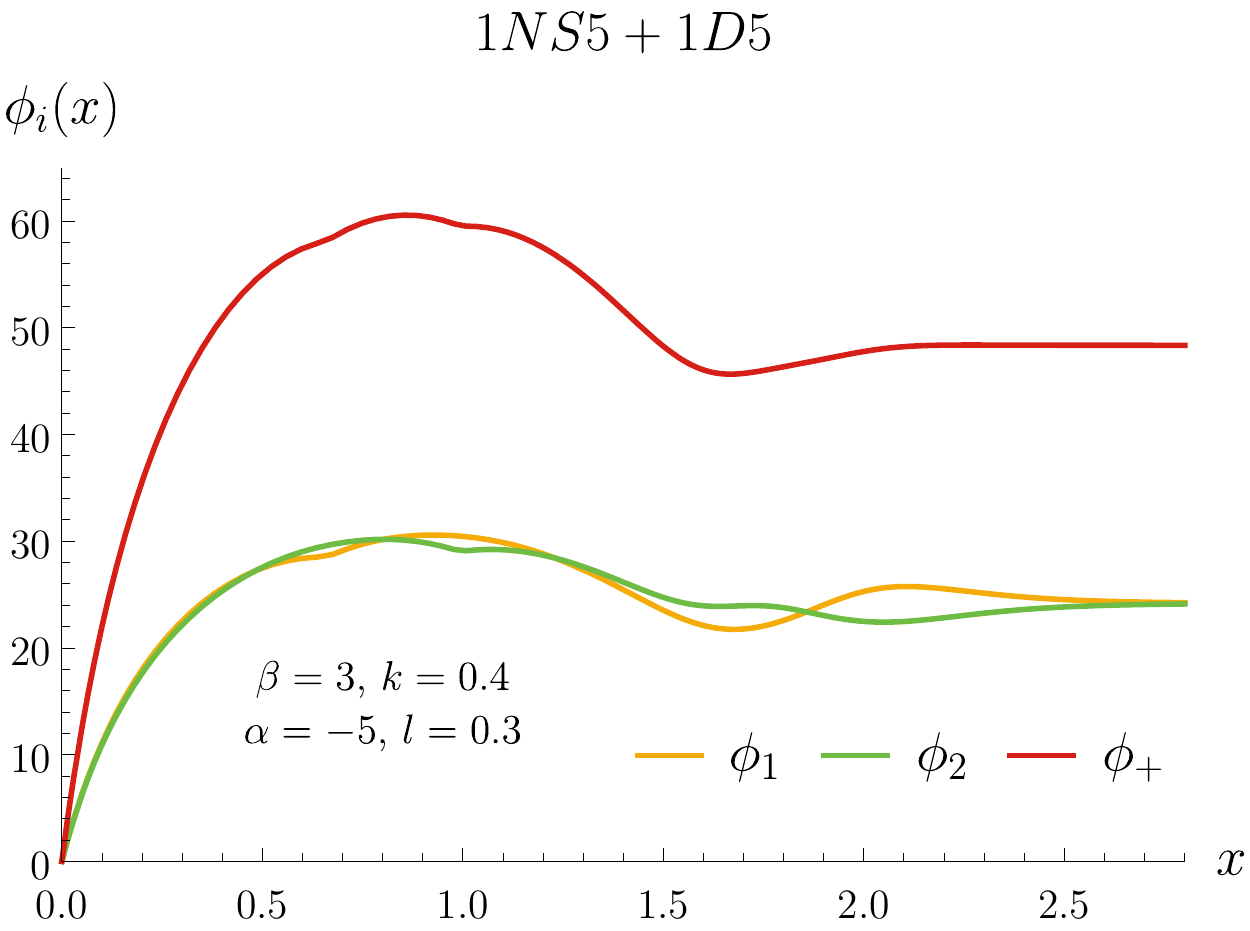}
\caption{\small The $\IS^5$ breathing mode in two illustrative examples of a single NS5-brane, and one NS5- and one D5-brane with different parameters (chosen as in Figure \ref{fig:different}).}
\label{fig:breathing}
\end{center}
\end{figure}

We consider the two scalars $\phi_1$, $\phi_2$ corresponding to the coefficients of the $ds_{\IS_1^2}$, $ds_{\IS_2^2}$ elements in the 10d metric (\ref{ansatz}). Hence their profiles in the solution are given by
\begin{equation}
     \phi_1\equiv f_1^2\ , \ \ \ \ \phi_2\equiv f_2^2
\end{equation}
Note that at large $r$ the asymptotic metric (\ref{the-asympt-ads}) implies that the (normalized) zero modes are
\begin{equation}
\phi_1^{(0)}(\varphi)=\frac{4}{\sqrt{3\pi}}\sin^2\varphi \quad , \quad \phi_{2}^{(0)}(\varphi)=\frac{4}{\sqrt{3\pi}}\cos^2\varphi
\label{zero-modes-phi12}
\end{equation}
The scalar $\phi_+$ parametrizing the size of the $\IS^5$ can thus be taken to correspond to the projection of $\phi_1+\phi_2$ in the solution along the zero mode, namely
\begin{equation}
\phi_+=\frac{4}{\sqrt{3\pi}}\int_{\pi/2}^\pi \left( \sin^2\varphi \,f_1^2 + \cos^2\varphi\, f_2^2\right)\, .
\label{breathing-5d}
\end{equation}
This cannot be solved analytically, but can be computed numerically. Some illustrative examples of the behaviour of $\phi_1$, $\phi_2$ and $\phi_+$ are shown in Figure \ref{fig:breathing}, for one NS5-brane and for one NS5- and one D5-brane. In the latter case, we have chosen different parameters for the two 5-branes, so as to see their different effect in $\phi_1$ and $\phi_2$.

As expected, the $\IS^5$ breathing mode goes to a constant value in the asymptotic AdS$_5$ region, and it goes to zero at the origin, where the $\IS^5$ shrinks and spacetime ends. It is easy to show that the scalar traverses an infinite distance in field space traversed in the finite spacetime distance to the origin; this leads to the metric singularity at the origin mentioned in the previous section. In fact, the behaviour of the 5d solution near the origin falls in the local description of dynamical cobordisms in \cite{Angius:2022aeq}, and in particular leads to the scaling relations of the spacetime distance $\Delta$, the field space distance ${\cal D}$ and the spacetime curvature scalar $R$, as follows. 

Near $r=0$ the dilaton is essentially constant, there is no RR 5-form flux, and the $\IS^5$ metric is round, the local description of the dynamical cobordism is just an $\IR^6$ split as $\IS^5$ slices shrinking as $r\to 0$. Indeed expanding (\ref{the-hs}) for small $r$ we have $h_1=4y$, $h_2=-4x$, which lead to the 10d metric
\beqa 
ds^2=F_1 \Big( F_2\, ds^2_{AdS_4} + dr^2 + r^2\big( d\varphi^2+\sin^2\varphi \,ds_{\IS_1^2}^2 + \cos^2\varphi\, ds_{\IS_2^2}^2\big)\Big)\, ,
\eeqa 
where $F_1$, $F_2$ are constants involving the brane parameters. We see that the $\IS^5$ is round, so that we recover a simpler reduction to get the 5d Einstein metric with a breathing mode 
\beqa 
ds^2= e^{\sqrt{\frac{5}{6}} \omega} ds_5^2+e^{-\sqrt{\frac{3}{10}}\omega}ds_{\IS^5}^2\, ,
\eeqa 
such that the 5d metric is in the Einstein frame and the scalar is canonically normalized. The spacetime distance to the singularity is
\beqa 
\Delta = \int g_{rr}^{1/2} dr\propto \int r^{5/3}dr\sim r^{8/3}
\eeqa 
The spacetime scalar curvature near $r=0$ is given by
\beqa 
R=-20 F_1^{-8/3}r^{-16/3}\quad \Rightarrow R\sim \Delta^{-2}
\eeqa 
On the other hand, the distance traversed in field space\footnote{One may worry that we are measuring the distance using the field variable $\omega$, whereas above we have used $\phi_+$ to describe the breathing mode. The result is however unchanged since near $r=0$ they are related by a field redefinition.} is
\beqa 
{\cal D}=\int g_{\omega\omega}^{1/2}d\omega=\omega\,\sim -\sqrt{\frac{10}{3}}\log r^2 \sim\,  -\sqrt{\frac{15}{8}}\log \Delta
\eeqa 
This reproduces the scaling laws in \cite{Angius:2022aeq} for a critical exponent $\delta=2\sqrt{\frac{8}{15}}$. Hence, the breathing mode attains infinite distance in field space at finite spacetime distance, in agreement for the criteria of a dynamical cobordism to nothing in \cite{Buratti:2021fiv,Angius:2022aeq}.

It is interesting to notice that the local description in \cite{Angius:2022aeq} captures only the collapse of the $\IS^5$, but not other features of the ETW brane, such as the KR bump. This shows that effects such as localization of gravity requires a description beyond that in \cite{Angius:2022aeq}. We will however show in Section \ref{sec:double-scaling} that it is possible to construct a local description rich enough to account for all features of the ETW brane, including localized gravity.

The behaviour of the $\IS^5$ breathing mode thus confirms that the 10d solutions describing ETW branes for AdS$_5\times\IS^5$ correspond to dynamical cobordisms from the 5d perspective. Note however that, although the behaviour near the origin is a local dynamical cobordism of the kind in \cite{Angius:2022aeq}, the whole ETW configurations involves additional structure, at a scale fixed by $\beta_1$, $\alpha_{g+1}$, necessary to remove the RR 5-form flux, and ultimately leading to localization of gravity. This is a novel feature compared with earlier realizations of dynamical cobordisms.

\begin{figure}[htb]
\begin{center}
\includegraphics[scale=.6]{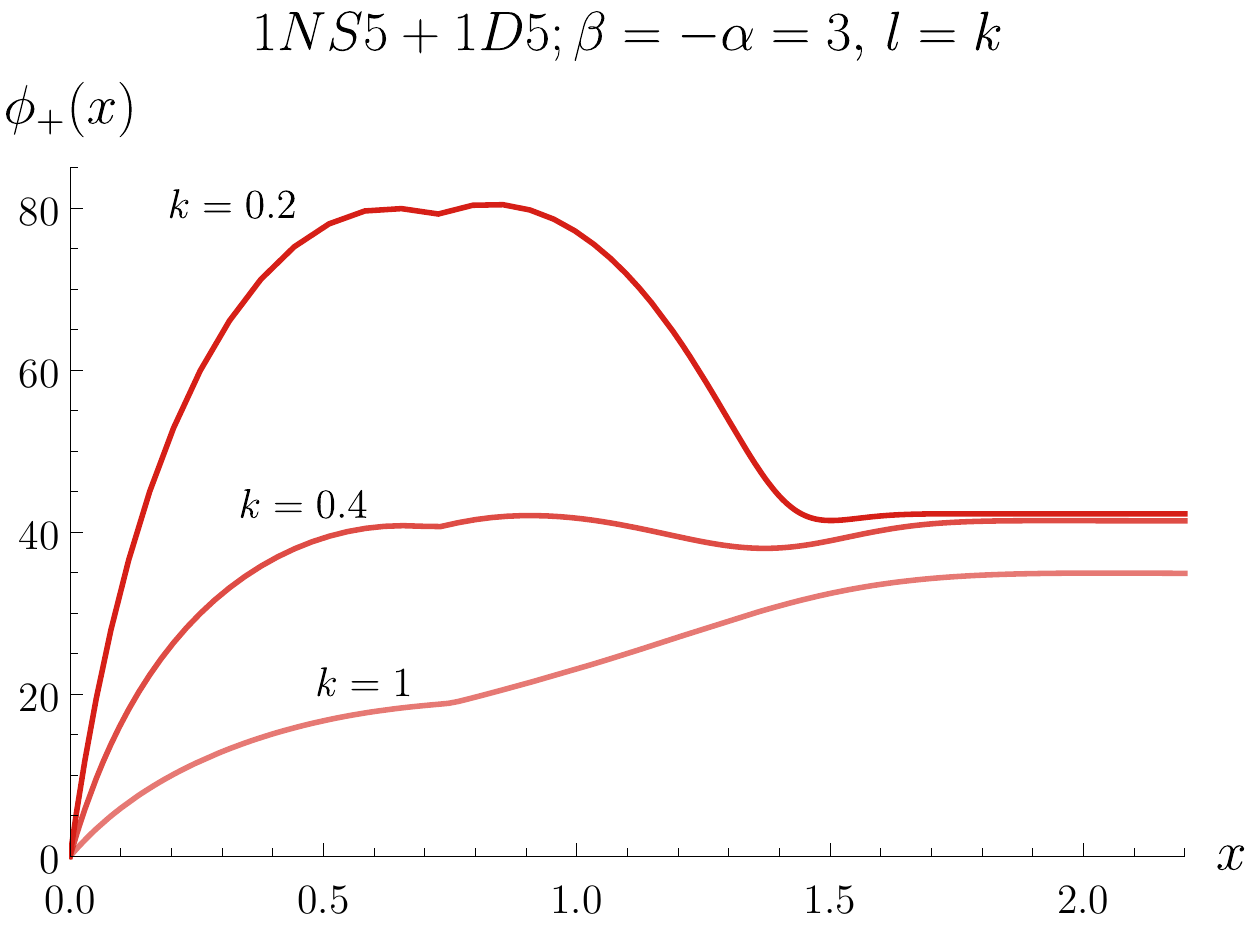}
\caption{\small The $\IS^5$ breathing mode for one NS5- and one D5-brane, for the choices of parameters in Figure \ref{fig:change-k}. As $k$ decreases, for fixed $\beta$, the size of the ETW region increases.}
\label{fig:neck}
\end{center}
\end{figure}

Another interesting feature is that, in the case with both NS5- and D5-branes the $\IS^5$ describes a bump with a maximum around the ETW brane region. It is easy to check numerically, see Figure \ref{fig:neck}, that the bump strengthens when $\beta\gg k^2$.
 This signals the growing of an almost independent compact geometry in the ETW region, with respect to which the asymptotic AdS$_5$ geometry is suppressed. This  will receive an intuitive explanation in Section \ref{sec:double-scaling}.

\subsubsection*{The $SO(3)\times SO(3)$ squashing}

The above discussion leads to a picture of a boundary described by a configuration corresponding to a KR-like brane in AdS$_5$, dressed with a running scalar dealing with the evolution and shrinking of the $\IS^5$, albeit in an $SO(6)$ invariant fashion. One may argue that this description misses important information about the boundary conditions, in particular the fact that they preserve only $SO(3)\times SO(3)$. It should be clear from our discussion above that it is easy to systematically include extra degrees of freedom, corresponding to higher KK modes in the reduction in the $\IS^1$ parametrized by $\varphi$. 
A general feature is that, since the $\IS^5$ tends to the round metric both at large $r$ and at small $r\simeq 0$, those 5d scalars will have a non-trivial profile only around the bump in the 5d metric; namely, they correspond to  worldvolume modes of the smoothed KR brane (see \cite{Kanda:2023zse} for KR branes with worldvolume scalars).

\begin{figure}[htb]
\begin{center}
\includegraphics[scale=.53]{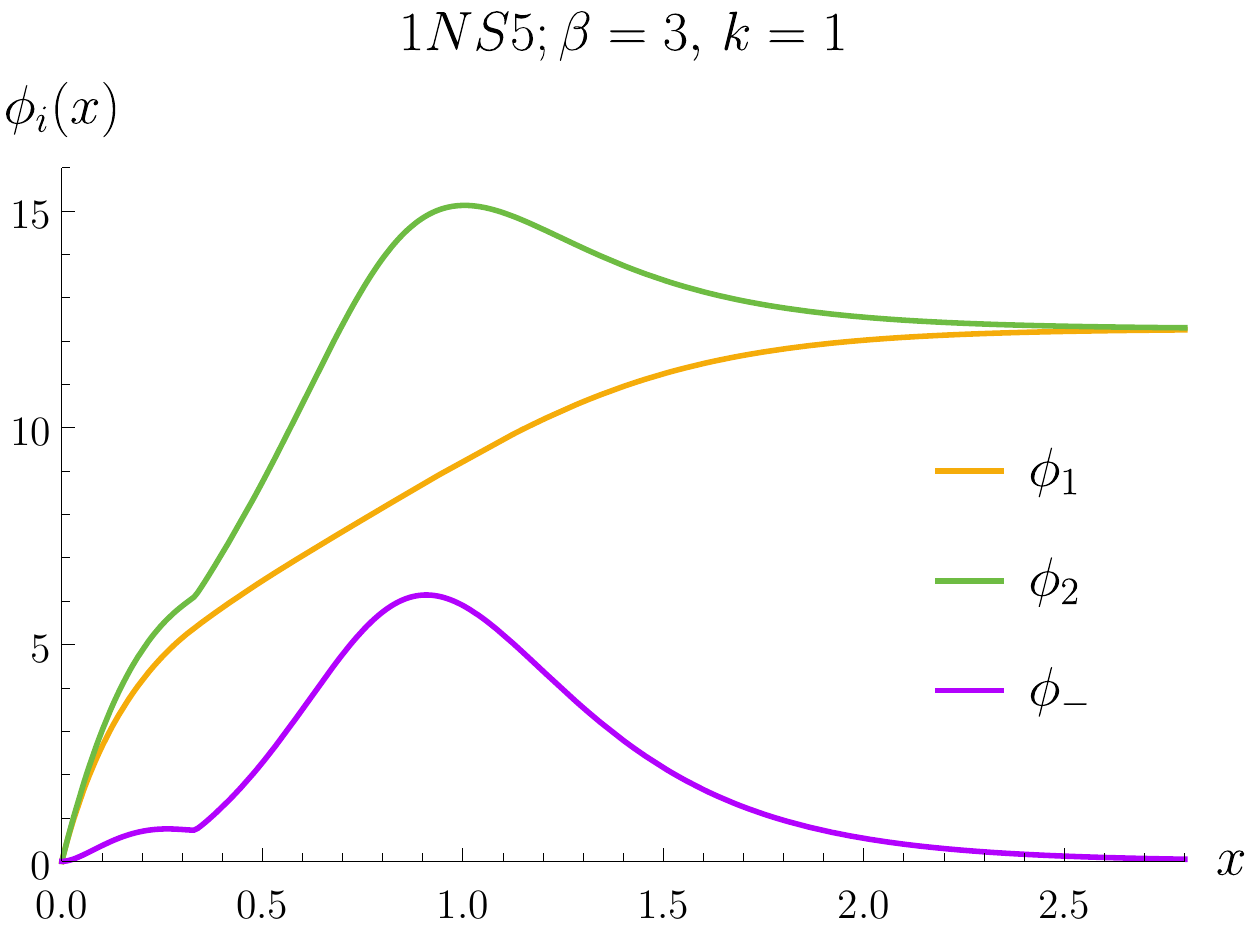}\hspace*{1cm}
\includegraphics[scale=.53]{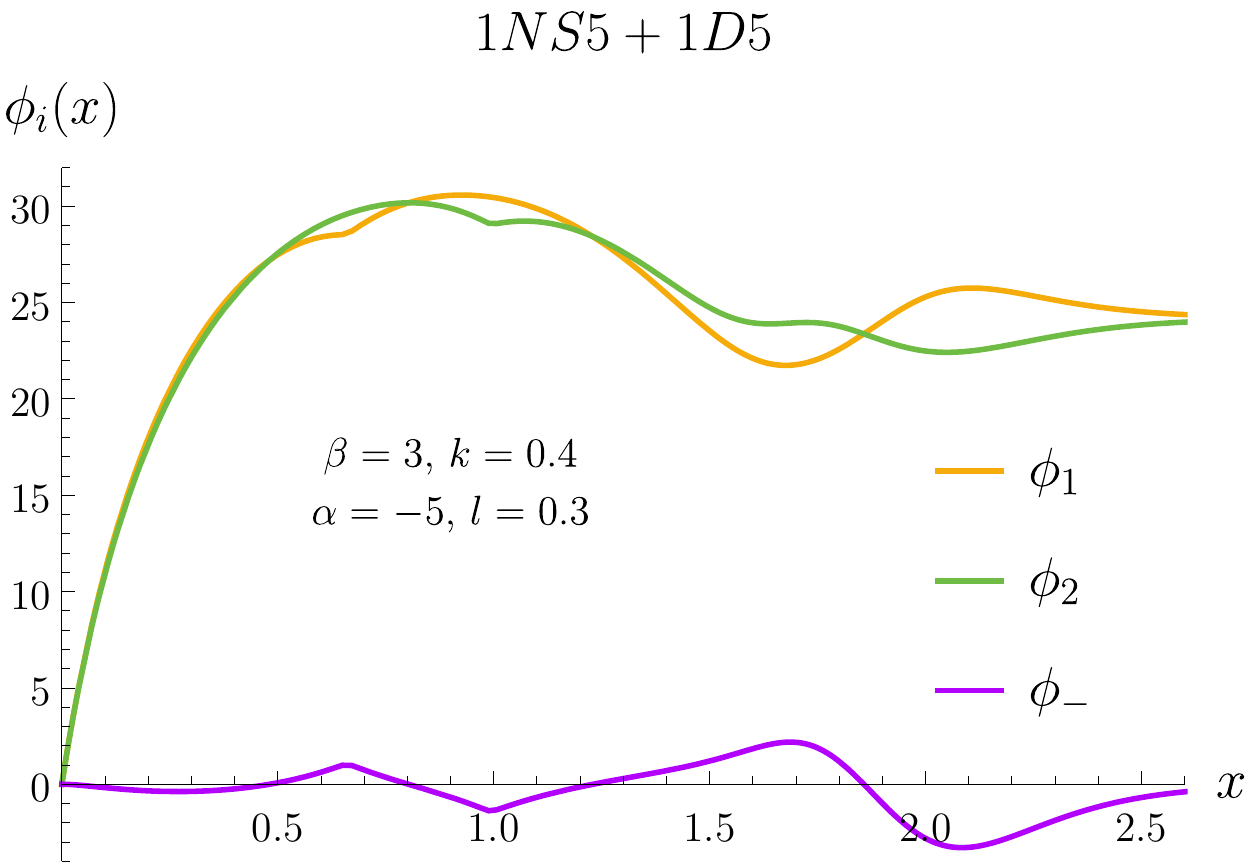}
\caption{\small The $SO(6)\to SO(3)\times SO(3)$ squashing mode, for the choice of 5-branes and parameters of Figure \ref{fig:breathing}.}
\label{fig:squashing}
\end{center}
\end{figure}

In particular, a simple  $SO(6)\to  SO(3)\times SO(3)$ breaking mode $\phi_-$ is obtained from the reduction of the 6d combination $\phi_-=\phi_1-\phi_2$ along the zero modes (\ref{zero-modes-phi12}), namely the antisymmetric combination of (\ref{breathing-5d}). Again, this integration is not amenable to analytic treatment, but can be studied numerically.
The resulting profiles in some illustrative examples are shown in Figures \ref{fig:squashing}.

As announced, the mode  $\phi_-$ is supported on the bump and goes to zero at large and small distance. It is interesting to notice that the scalar associated to the corresponding 5-brane (namely, $\phi_1$ for D5-branes and $\phi_2$ for NS5-branes) suffers a discontinuity in crossing the 5-brane location. This implies that the discontinuities in $\phi_-$ have opposite sign for NS5- and D5-branes, providing a 5d criterion to distinguish them. 

As a last remark, we would like to emphasize that, besides the kind of KK modes of the metric discussed above, there are additional scalars arising from the reduction of the 2-forms, but we skip their discussion (see \cite{Aharony:2011yc} for a discussion of the lowest such mode). In fact, there is actually a tower of $SO(6)\to SO(3)\times SO(3)$ breaking modes localized on the ETW branes, with no mass hierarchy among them. This is actually expected, as these modes in AdS$_4$ are the gravity duals of the chiral operators in the holographic BCFT. In fact, this latter language may provide the best approach to characterizing the ETW brane worldvolume dynamics \cite{Assel:2011xz}, as we also discuss in the next Section. 

\section{Zoom into the ETW brane: A double scaling limit}
\label{sec:double-scaling}

In this section we present a, to our knowledge new, scaling limit of the 10d solutions in section \ref{sec:case-of-one}, which will be useful to isolate the physics of the ETW brane configuration ending the AdS$_5$ spacetime. It allows for a simple intuitive understanding of some key properties of the ETW branes, noticed in earlier sections. Interestingly, the limit ends up being equivalent to closing off the last AdS$_5\times\IS^5$ asymptotic region in the solutions in section \ref{sec:case-of-one}, and is closely related to solutions considered in \cite{Assel:2011xz}, and setups of wedge holography \cite{Akal:2020wfl,VanRaamsdonk:2021duo}.

\subsection{A scaling symmetry}
\label{sec:scaling}

Let us start by considering an invariance property of the 10d solutions in section \ref{sec:case-of-one}, focusing on the case of one asymptotic AdS$_5\times \IS^5$ region. Consider the scaling
\beqa
w\to \lambda w\quad ,\quad k_a\to \lambda k_a\quad ,\quad  l_b\to \lambda l_b \quad ,\quad  \beta_a\to \lambda^2 \beta_a \quad ,\quad \alpha_b\to \lambda^2 \alpha_b\, ,
\label{limit-conf}
\eeqa
which preserves the ordering (\ref{the-order-bis}). From (\ref{the-ds}) and (\ref{the-hs}), we have the scaling
\beqa
d_a\to \lambda d_a \quad ,\quad {\tilde d}_b\to \lambda {\tilde d}_a  \quad ,\quad h_1\to \lambda h_1\quad ,\quad h_2\to\lambda h_2\, .
\eeqa
Then from (\ref{wnn}) and (\ref{dilaton}), we get
\beqa
W\to W\quad ,\quad N_1\to \lambda^2 N_1\quad ,\quad N_2\to \lambda^2 N_2\quad e^{2\Phi}\to e^{2\Phi}\, ,
\eeqa 
and the scalings of the functions (\ref{the-fs}) are
\beqa
\rho^2\to \lambda^{-1}\rho^2\quad ,\quad f_1^2\to \lambda f_1^2\quad ,\quad f_2^2\to \lambda f_2^2\quad ,\quad f_4^2\to \lambda f_4^2
\eeqa
Hence, using also (\ref{2dmetric}), the 10d metric (\ref{ansatz}) experiences an overall rescaling
\beqa
ds^2\to \lambda ds^2\, .
\eeqa
The metric is thus invariant up to an overall constant rescaling.

\subsection{Double scale zooming into the ETW configuration}
\label{sec:zoom}

Our motivation in this work is to better understand the structure of the ETW configuration ending spacetime. For that purpose, we now consider a double scaling limit which zooms into the ETW configuration. This will be an interesting approach to isolate its properties, and to provide a direct 10d description of the KR AdS$_4$ branes.

A naive limit to focus on the region ending spacetime would be to just zoom onto $w=0$. However, this is only a smooth point in the 10d geometry, which misses all the information about the 5-brane configurations, hence leads to a rather trivial limit. Therefore we consider a more interesting double scaling limit
\beqa
w\to \lambda w\quad ,\quad k_a\to \lambda k_a\quad ,\quad  l_b\to \lambda l_b \quad ,\quad {\rm as}\; \lambda\to 0\, ,
\label{the-limit}
\eeqa
which sends all the 5-branes towards the origin as we zoom onto it (and them). In order to preserve the ordering (\ref{the-order-bis}), we also take
\beqa
\beta_a\to \lambda^2\beta_a \;\; {\rm for}\; a\neq 1\quad ;\quad 
\alpha_b\to \lambda^2 \alpha_b \;\; {\rm for}\; b\neq g+1\, .
\label{the-limit-bis}
\eeqa
Namely, we send all points in the ordering to the origin, except for the extreme ones $\beta_1$, $\alpha_{g+1}$. We emphasize that this scaling is possible only if the configuration contains both NS5- {\em and} D5-branes simultaneously. In the absence of D5-branes (i.e. no $l_b$'s), the sequence (\ref{the-order-bis}) contains no $\alpha$'s (equivalently, the only one in (\ref{the-order}) had already been sent to $0$ in closing off the AdS$_5\times\IS^5$ region at the origin, so it cannot kept finite in the scaling limit). We will come back to the need of both kinds of 5-branes later on.

The reason to keep $\beta_1$, $\alpha_{g+1}$ finite is that, as explained above, they set the effective size of the fat ETW brane. Note also that this is the crucial difference with respect to the scaling (\ref{limit-conf}) above, and will lead to a non-trivial change in the configuration.

In particular, the leading terms in (\ref{the-ds}) are
\beqa
d_1=\lambda^{-1}\frac{\beta_1}{2k_1} \prod_{c\neq 1}^n\frac{(k_1^2-\beta_c)}{(k_1^2-k_c^2)}\quad ,\quad
 d_a=-\lambda^{-1}\frac{\beta_1(\beta_a-k_a^2)}{2k_a(k_a^2-k_1^2)} \prod_{c\neq a,1}^n\frac{k_a^2-\beta_c}{k_a^2-k_c^2}\quad\, \; a\neq 1\quad\quad\quad\quad\quad\quad\quad
 \label{the-ds-limit} \\
\!\!{\tilde d}_{g-n}\!=\!-\lambda^{-1}\frac{\alpha_{g+1}}{2l_{g-n}} \prod_{c\neq g-n}^m \!\!\!\! \frac{(l_{g-n}^2+\alpha_{c+n+1})}{(l_{g-n}^2-l_c^2)} ,\;
{\tilde d}_b\!=\!\lambda^{-1}\frac{\alpha_{g+1}(-\alpha_{b+n+a}-l_b^2)}{2l_b(l_b^2-l_{g-n}^2)}\!\!\! \prod_{c\neq b,g-n}^n \!\! \!\!\frac{(l_b^2-\alpha_{c+n+1})}{(l_b^2-l_c^2)}\; ,\quad\quad\quad\quad\nonumber\\ b\neq g-n\, .\quad\quad\quad\quad \nonumber
\eeqa
Since they scale as $\lambda^{-1}$, the first term in (\ref{the-hs}) is subleading and drops out, leading to
\beqa
h_1=\;2\sum_{b=1}^m{\tilde d}_b \log\left(\frac{|w+il_b|^2}{|w-il_b|^2}\right)\quad ,\quad
h_2=-2\sum_{a=1}^n d_a \log\left(\frac{|w+k_a|^2}{|w-k_a|^2}\right)\, ,
\label{the-hs-limit}
\eeqa
They also scale as $\lambda^{-1}$, and using (\ref{wnn}) we get that $W$ scales as $\lambda^{-4}$, and $N_1$, $N_2$ scale as $\lambda^{-6}$. From (\ref{dilaton}), the dilaton has no scaling with $\lambda$, and from (\ref{the-fs}), we have that $\rho^2$ scales as $\lambda^{-3}$, while $f_1^2$, $f_2^2$ and $f_4^4$ all scale as $\lambda^{-1}$. Note then that the 10d metric (\ref{ansatz}) scales as $\lambda^{-1}$. Forgetting this overall factor, we have a metric of the kind considered in section \ref{sec:case-of-one}, with the modified $h$-functions in (\ref{the-hs-limit}).

Let us discuss some  key properties of the resulting geometry: The asymptotic AdS$_5\times \IS^5$ region at $r\to\infty$ disappears. Correspondingly, the asymptotic 5-form flux effectively disappears. The simplest way to show this is that
the point $r\to \infty$ is actually at finite distance in the 10d Einstein metric, and in fact corresponds to a smooth point in the geometry. In other words, the 6d geometry $\IX_6$ transverse to AdS$_4$ is actually compact: the configuration is a compactification AdS$_4\times \IX_6$. 

This may seem surprising, but actually matches the intuition that the scaling limit encodes the dynamics of the ETW brane configuration. In particular, its compactness implies the existence of a localized graviton, one of the main properties of our ETW branes. Actually the localized graviton is a massive one in the full geometry before the scaling limit, due to its coupling to the bulk AdS$_5$; this is a subleading $\lambda^{-1}$ effect in our description and disappears in the limit. The fact that the massive graviton can become massless in a continuous fashion is related to the discussion in \cite{Karch:2001jb}.

There is also an extra bonus: the fact that the scaling limit (and its graviton zero mode) exist only if both NS5- and D5-are present explains the observation in section \ref{sec:reduct} that the KR bump localizing gravity in the 5d metric appears only when both kinds of 5-branes are present. Note also that an additional requirement for localized gravity in section \ref{sec:reduct} was the hierarchy between $\alpha,\beta$ and the 5-brane locations $k,l$; this is automatically built in in the scaling limit, since the former are kept finite while the latter are sent to 0.

The fact that the solution has only 5-brane charge and no 5-form flux would seem to suggest that it is just the near horizon solution of a stack of 5-branes. However, our solution maintains all the information about the $k_a$, $l_b$, $d_a$, ${\tilde d}_b$, hence all the information about the boundary conditions of the holographic CFT. The limit thus isolates the relevant information about the ETW brane in the gravity side.

In fact, 10d geometries defined by logarithmic functions $h_1, h_2$ as in (\ref{the-hs-limit}) were considered in \cite{Assel:2011xz,VanRaamsdonk:2021duo} to define the gravitational dual of 3d $\NN=4$ theories. In our context, this AdS$_4$/CFT$_3$ duality is just the holography of the gravitational theory on the ETW brane worldvolume and the BCFT of the initial 4d CFT.

\subsection{The limit revisited: closing off AdS$_5\times\IS^5$}
\label{sec:revisit}

We now provide an alternative description for the limit (\ref{the-limit}), (\ref{the-limit-bis}). Recall the invariance at the beginning of this section i.e. the rescaling (\ref{limit-conf}) of $w$ and all the special points $\alpha_b$, $\beta_a$, $k_a^2$, $l_b^2$. Using this invariance, the above limit of sending to zero $w$ and all the special points, except $\alpha_{g+1}$, $\beta_1$, is equivalent to keeping $w$ and all special points fixed, except for $\alpha_{g+1}$, $\beta_1$, which are sent to infinity. This can be seen from (\ref{the-ds-limit}), by reinterpreting the prefactor $\lambda^{-1}$ as a scaling associated to $\alpha_{g+1}$, $\beta_1$, while the other quantities in the expressions remain constant. Note that, as above, the limit exists only if both NS5- and D5-branes are present in the configuration, otherwise at least one of the parameters  $\alpha_{g+1}$, $\beta_1$ is missing, and we cannot send it to infinity.

In this description, many of the features observed in the previous section have a simple interpretation. Recalling the description in section \ref{sec:bagpipes}, asymptotic AdS$_5\times\IS^5$ regions arise `bare' branch points $e_i$, and they are closed off when the adjacent $\alpha$ and $\beta$ collapse onto it. In this description, the single asymptotic AdS$_5\times\IS^5$ region in the solution in section \ref{sec:case-of-one} is associated to the branch point $e_{2g+2}$, which is located at infinity.  The above limit of sending $\beta_1$ and $\alpha_{g+1}$ to infinity simply corresponds to collapsing them onto $e_{2g+2}$, and therefore to closing off the AdS$_5\times\IS^5$ region. 

Since this was the only non-compact spike of the initial geometry, the resulting space is compact. It describes the gravitational dual of the BCFT defined by the 5-brane configuration. Since it corresponds to a compactification, it has a graviton zero mode; this nicely corresponds to the graviton localized by the ETW brane, matching the expected behaviour for AdS$_4$ KR branes.
Again, the coupling of the ETW brane to the to the asymptotic AdS$_5$ region, and the corresponding mass for the localized graviton, are switched off in the limit, leaving the graviton massless. Note also that we recover the connection between having both NS5- and D5-branes simultaneously (and the hierarchy of the parameters $\alpha_{g+1},\beta_2$  and $k_a,l_b$) and the appearance of localized gravity, observed in section \ref{sec:reduct}, as in the previous section.

\subsection{Reduction in the double scaled limit}
\label{sec:reduct-limit}

We have noticed some relations between the localization of gravity in the 5d effective action approach in section \ref{sec:reduct} and in the scaling limit. This motivates performing a 5d reduction of the 10d solution obtained after the scaling limit, namely using the functions (\ref{the-hs-limit}). This is very analogous to the procedure in section \ref{sec:eft}, hence we simply sketch some general features, in a few plots analogous to those in section \ref{sec:eft}. Note that, due to the compactness of the internal space, the plots have finite extent in the radial direction.

\begin{figure}[htb]
\begin{center}
\includegraphics[scale=.53]{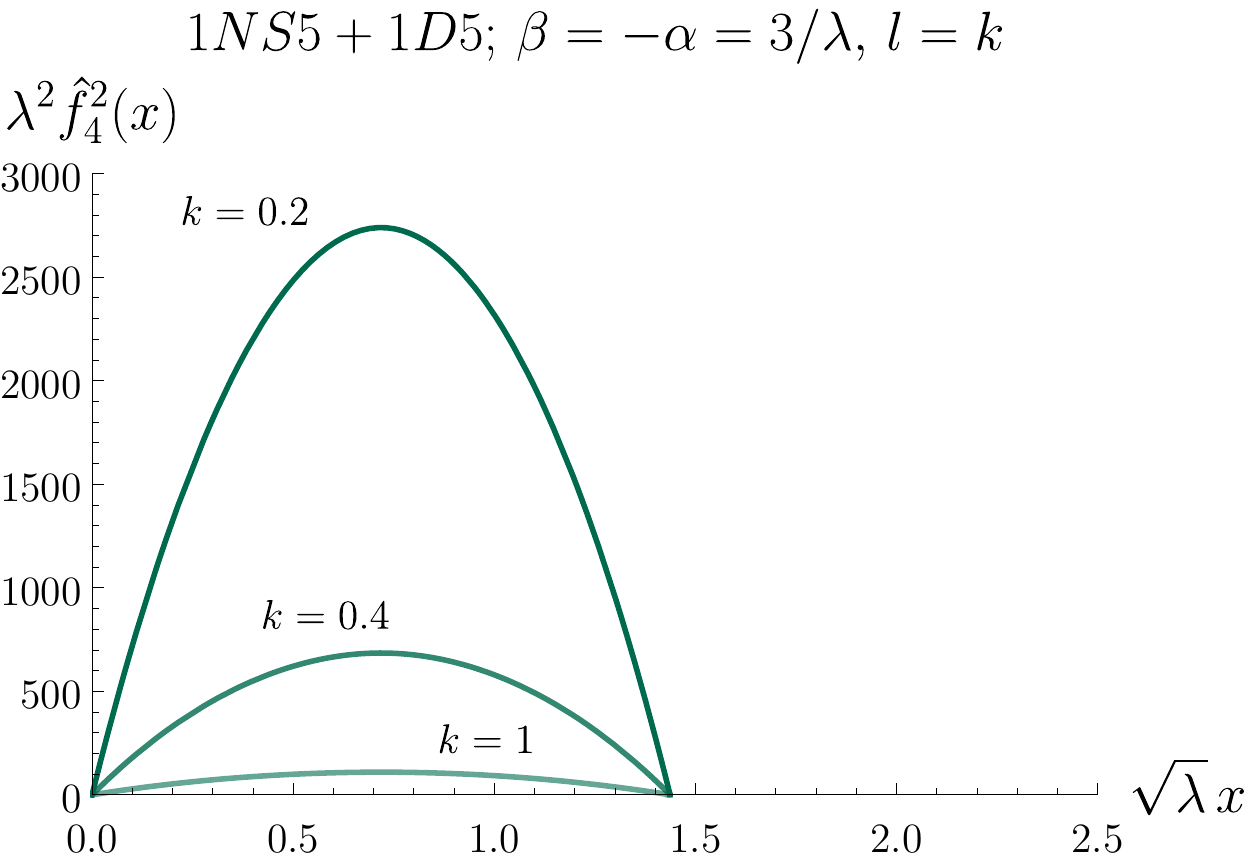}\hspace*{1cm}
\includegraphics[scale=.53]{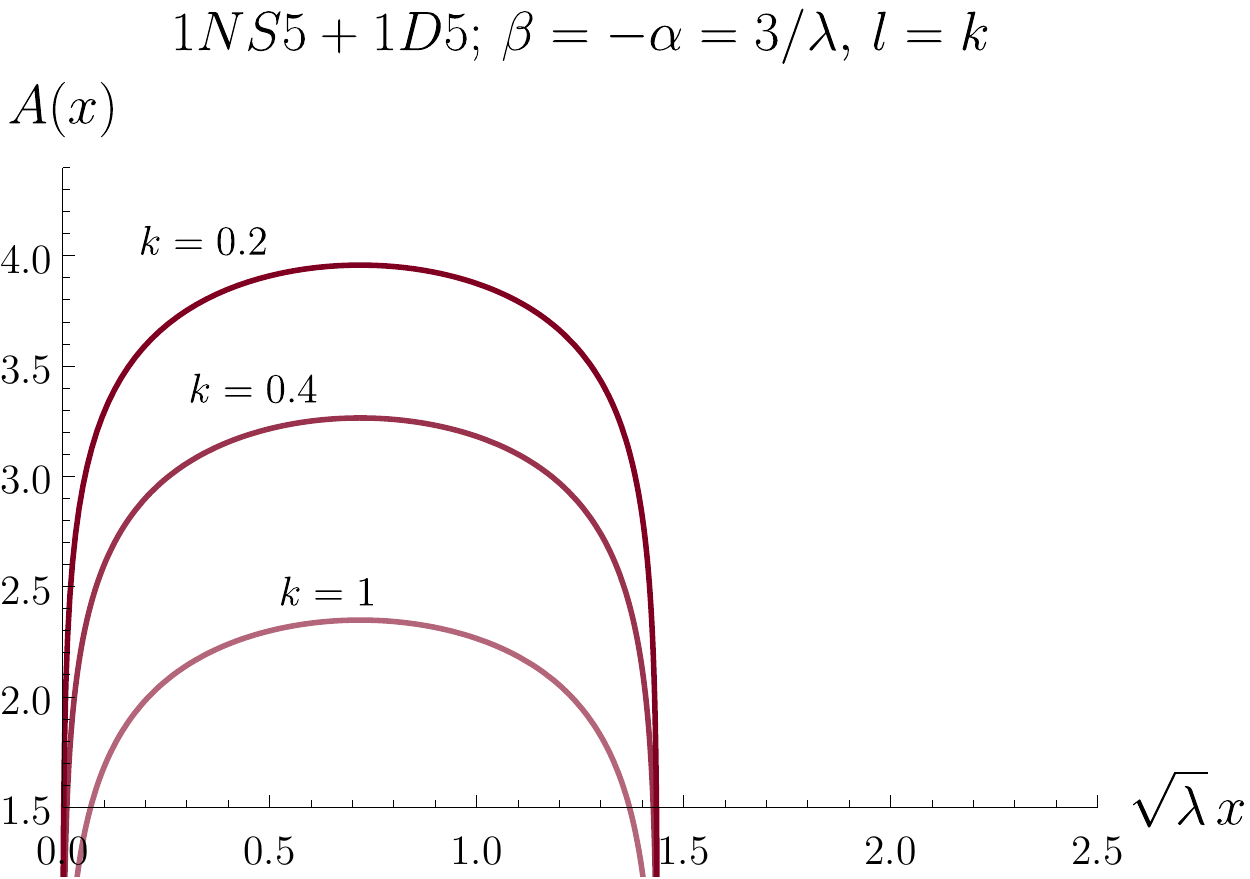}
\caption{\small Warp factors in the ETW brane for one NS5- and one D5-brane, for a choice of parameters related to those in Figure \ref{fig:change-k}.
}
\label{fig:warp-double-scaling}
\end{center}
\end{figure}

Figure \ref{fig:warp-double-scaling} provides the warp factors of the metric in the scaling limit, for the choices of brane configurations and parameters of Figure \ref{fig:change-k}. The axes scale with $\lambda$ as dictated by the KR metric \eqref{5d_metric_KR}. It is clear that the solution after the scaling provides a very good description of the ETW brane region of the full solution in Figure  \ref{fig:change-k}. 
Note that $\hat{f}_4^2$ goes to zero (and $A\to -\infty$) at the endpoints of the finite extent of the geometry, signalling the closing off of the asymptotic AdS$_5$ region. \\

\begin{figure}[htb]
\begin{center}
\includegraphics[scale=.53]{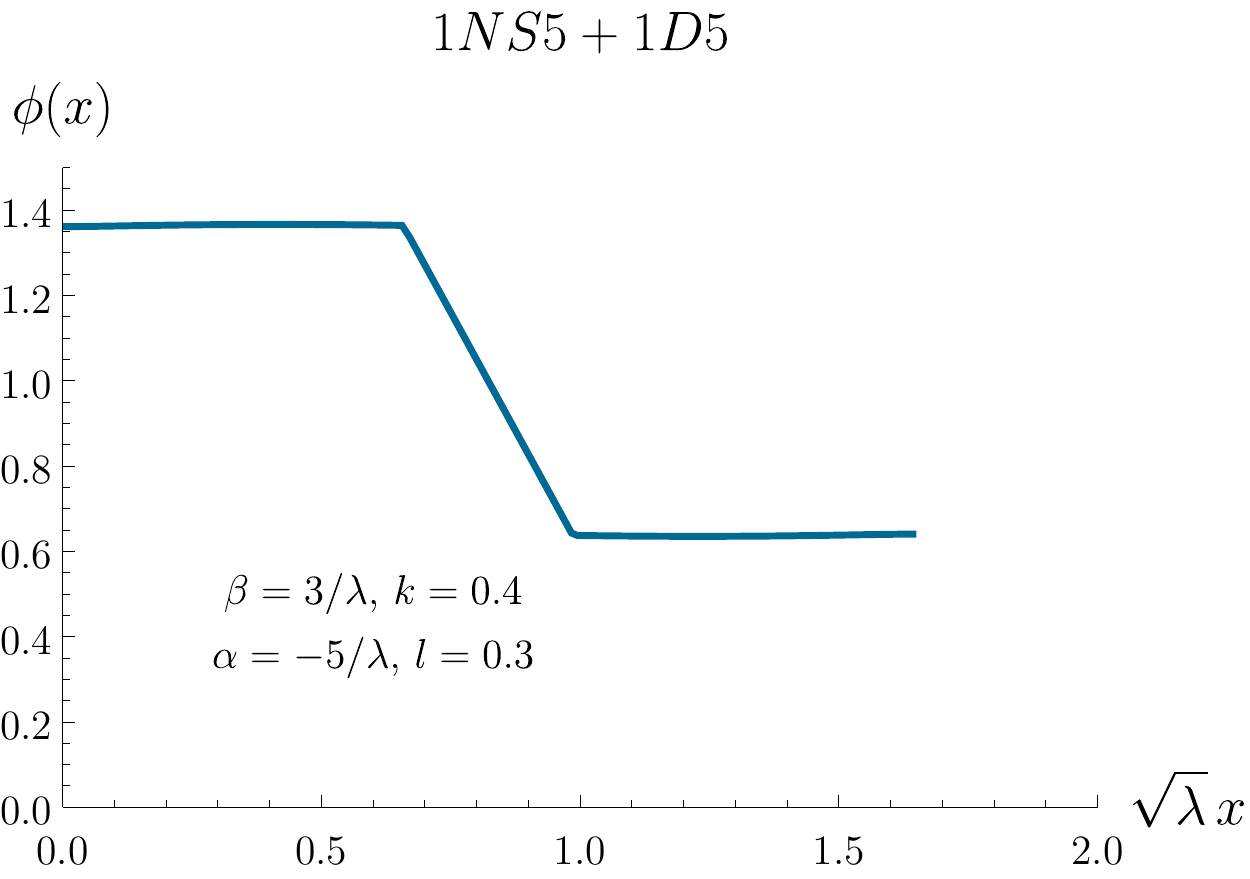}\hspace*{1cm}
\includegraphics[scale=.53]{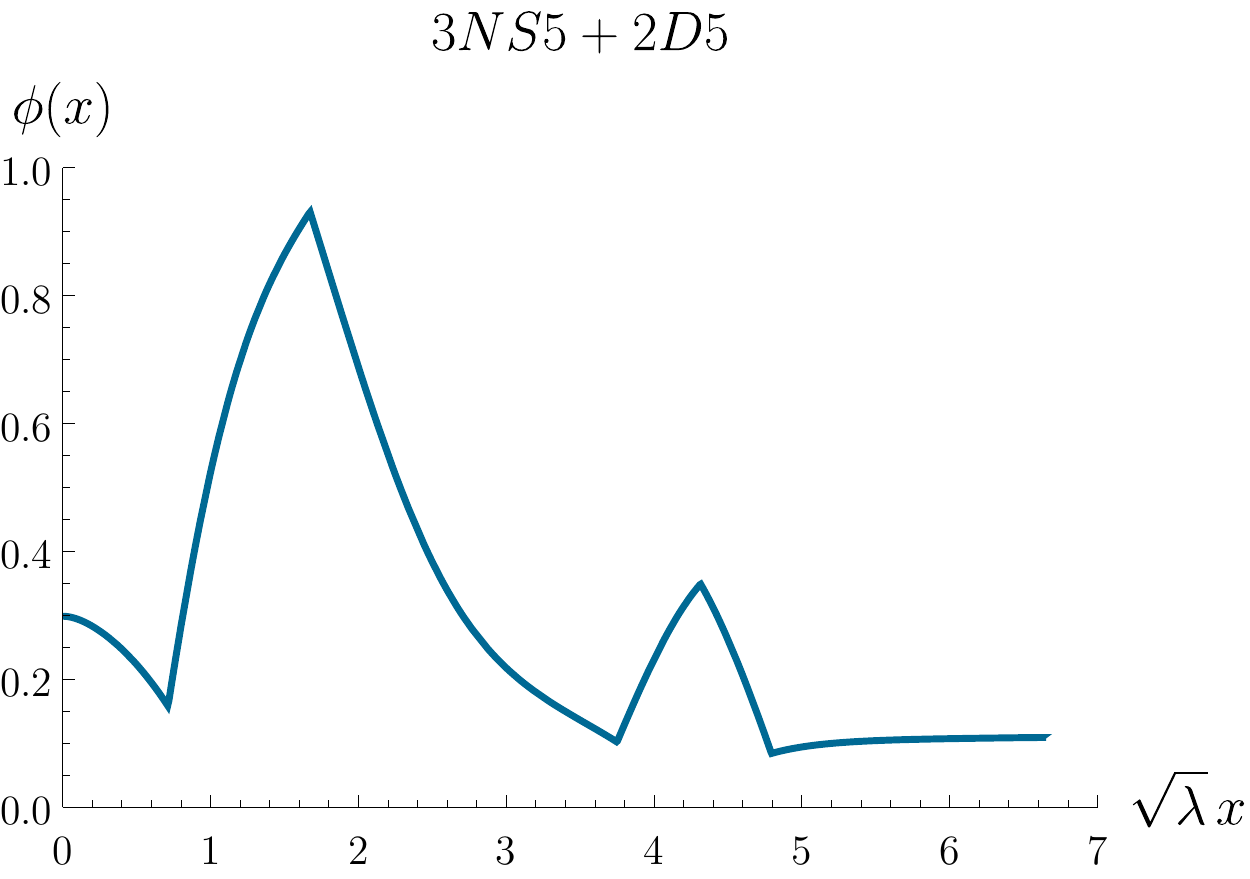}
\caption{\small Dilaton in the ETW brane for one NS5- and one D5-brane, for the choice of parameters of Figure \ref{fig:dilaton-one} and Figure \ref{fig:dilaton-many}-right.
}
\label{fig:dilaton-double-scaled}
\end{center}
\end{figure}

The behaviour of the dilaton is shown in two prototypical cases in Figure \ref{fig:dilaton-double-scaled}. Since there is no AdS$_5$ asymptotic regions, there is an undetermined constant (the average of the dilaton in the ETW volume) which we have fixed to some arbitrary value.

\begin{figure}[htb]
\begin{center}
\includegraphics[scale=.53]{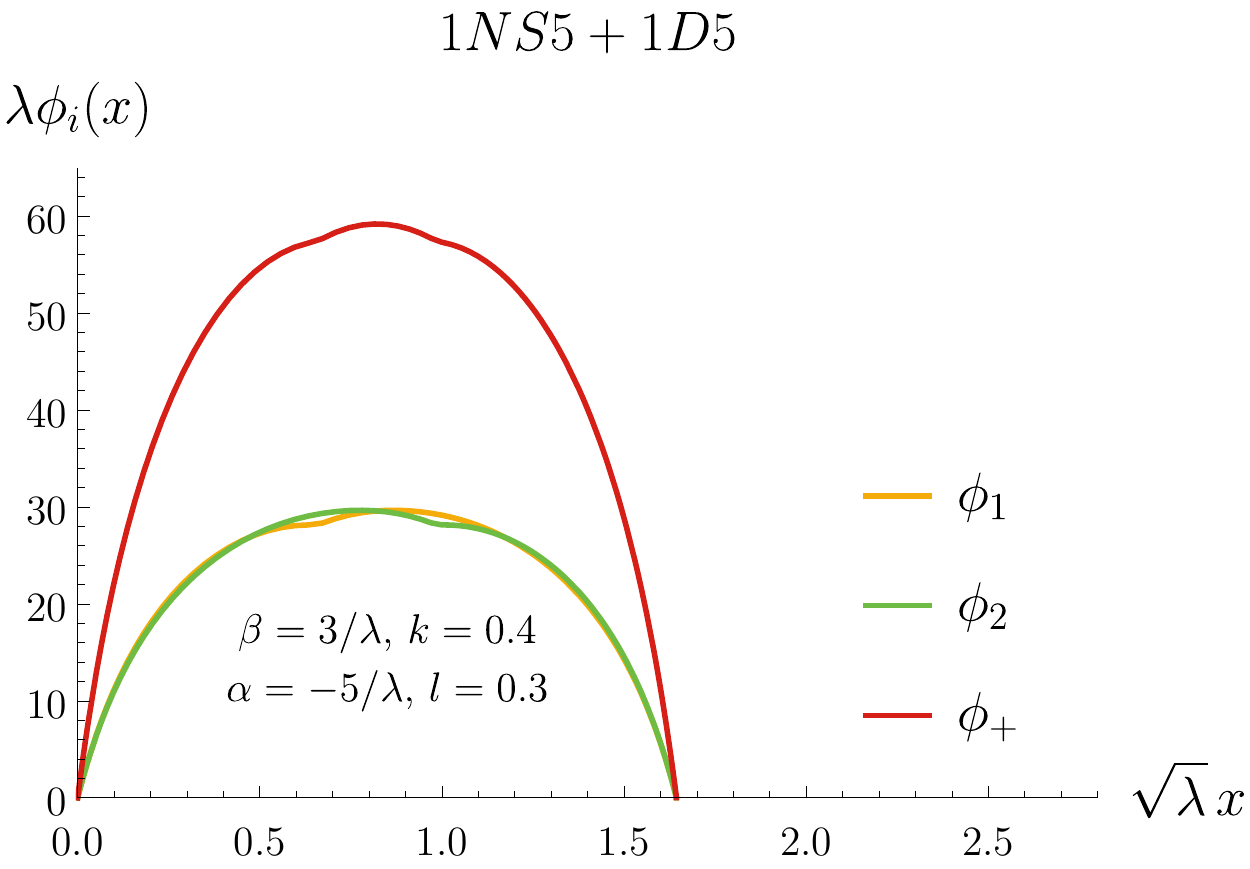}\hspace*{1cm}
\includegraphics[scale=.53]{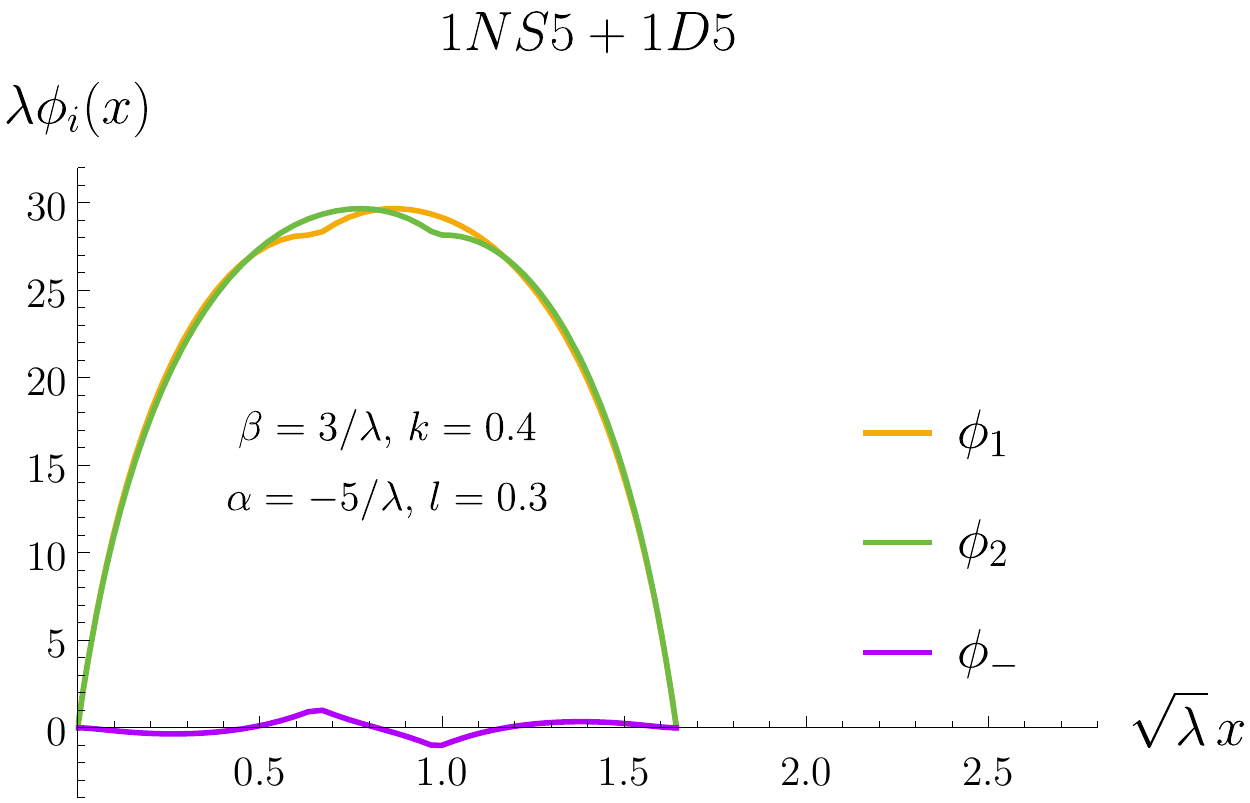}
\caption{\small The scalars $\phi_\pm$ for one NS5- and one D5-brane, for the choice of parameters of Figure \ref{fig:warp-double-scaling}.}
\label{fig:scalars-double-scaled}
\end{center}
\end{figure}

One can also study the scalars $\phi_+$, describing the evolution of the $\IS^5$ size, and $\phi_-$, describing its squashing. In Figure \ref{fig:scalars-double-scaled} we show them in an illustrative example. Note that the ETW geometry has $\IS^6$ topology, with the 5-branes introducing deviations from the round metric.

\medskip

In the above discussion we have worked in the 5d perspective and studied how different quantities evolve in the direction $r$. Since the latter is compact, it also makes sense to perform a dimensional reduction on it and describe the dynamics on AdS$_4$. This is actually in the spirit of AdS$4$/CFT$_3$ already considered in \cite{Assel:2011xz}, so we will not develop it further in this work.

\section{ETW branes in gravity duals of 4d $\NN=2$ orbifold theories}
\label{sec:n2}

In holography, understanding a gauge/gravity dual pair leads to many other examples with reduced (super)symmetry by the procedure of orbifolding/orientifolding. In this and the following sections we will describe quotients which are compatible with structure of the BCFT, and which hence provide ETW configurations for AdS$_5$ geometries with orbifolded internal spaces.

In this section we consider D3-branes at perturbative orbifold singularities $\IC^2/\IZ_k$, projecting the 4d $\NN=4$ SYM theory down to a 4d $\NN=2$ quiver gauge theory \cite{Douglas:1996sw}, whose gravity duals are AdS$_5\times\IS^5/\IZ_k$ geometries.
We will show that the configurations of NS5- and D5-branes of the parent theory are invariant under the orbifold action an descend to the orbifold quotient. Hence, they define BCFTs for the 4d $\NN=2$ quiver gauge theories, and their gravity dual are suitable orbifolds of the ETW configurations in section \ref{sec:case-of-one}.

\subsection{The 4d $\NN=2$ quiver gauge theory and its gravity dual}
\label{sec:quiver}

Consider a set of D3-branes located at the origin in a $\IC^2/\IZ_k$ singularity\footnote{One may consider $\IC^2/\Gamma$ singularities, with $\Gamma$ other subgroups of $SU(2)$ (i.e. $D_k$, $E_{6,7,8}$). We leave this development for future work.}. For concreteness, we let the D3-branes span the direction 0123, and let the generator $\Theta$ of $\IZ_k$ act on the complex coordinates $z_1=x^4+ix^5$, $z_2=x^7+ix^8$ as
\beqa
\Theta: (z_1,z_2)\;\rightarrow\; (e^{\frac{2\pi i}k z_1}, e^{-\frac{2\pi i}{k}z_2})\, .
\label{zk-generator}
\eeqa
One must also specify the action of $\Theta$ on the Chan-Paton indices of the D3-branes. Considering a stack of $Nk$ D3-branes in the covering space, the orbifold Chan-Paton action is a $U(Nk)$ matrix $\gamma_{\Theta,3}$ of order $k$, in other words, a representation of dimension $Nk$ of $\IZ_k$.  
In order to obtain 4d $\NN=2$ SCFT's \cite{Kachru:1998ys,Lawrence:1998ja,Hanany:1998ru,Hanany:1998it}, we choose $N$ copies of the regular representation
\beqa
\gamma_{\Theta,3}=\id_N\otimes (\,1,e^{2\pi i/k}, \ldots,e^{2\pi i\, (k-1)/k}\,)\, .
\label{cps}
\eeqa
The theory on the worldvolume of the D3-branes has gauge group\footnote{ The $U(1)$'s are massive due to BF couplings on the D3-brane worldvolumes; this is dual to the brane bending phenomenon in \cite{Witten:1997sc}.} $\prod_i SU(N)_i$ and hypermultiplets in the bifundamental representation $(\fund_i,\antifund_{i+1})$. This is usually represented as a circular quiver with $k$ nodes, and links joining neighbouring nodes.

As discussed in \cite{Kachru:1998ys,Lawrence:1998ja,Hanany:1998it}, the gravity dual of these theories is given by type IIB on AdS$_5\times \IS^5/\IZ_k$. The action of $\IZ_k$ is as follows: Introduce and extra complex coordinate $z_3=x^6+ix^9$, and describe  the $\IS^5$ as
\beqa
|z_1|^2+|z_2|^2+|z_3|^2=R^2\, .
\eeqa
The action of $\IZ_k$ on $\IS^5$ leaves $z_3$ invariant and acts on $z_1,z_2$ as in (\ref{zk-generator}). There is an $\IS^1$ of $\IC^2/\IZ_k$ fixed points corresponding to $z_1=z_2=0$, $|z_3|=R$. This leads to a set of fields localized at such singularities, see \cite{Hanany:1998it,Gukov:1998kk} for their holographic dictionary.

\subsection{The ETW orbifold configurations}
\label{sec:etw-n2}

In the brane construction it is clear that we can introduce NS5- and D5-brane boundary defects in the parent $\NN=4$ theory which survive the orbifold projection and hence yield boundary defects in the daughter $\NN=2$ theories. In fact, as in Section \ref{sec:holobound}, we introduce NS5-branes along 012456 (hence, spanning $z_1$, localized at the origin in $z_2$, and along the real axis in $z_3$) and D5-branes along 012789 (hence, at the origin in $z_1$, spanning $z_2$, and along the imaginary axis in $z_3$).

The action of $\IZ_k$ on D5-branes is easily discussed, in close analogy with D3-branes (the action on NS5-branes\footnote{Orbifolds of brane constructions with NS5-branes and $\IC^2/\IZ_k$ orbifolds were considered in a closely related setup with D4-branes in \cite{Lykken:1997gy}, which are a T-dual realization of brane boxes \cite{Hanany:1997tb,Hanany:1998ru,Hanany:1998it}.} should also be similar, due to S-duality of the underlying $\NN=2$ theory). Skipping a detailed derivation, it suffices to say that the orbifold action on 5-brane gauge degrees of freedom splits them into $k$ stacks, which are also organized into a cyclic quiver. The $\IZ_k$ quantum symmetry of the orbifold theory (simultaneous $\IZ_k$ rotations of the cyclic quivers of D3- and 5-branes), implies that the $i^{th}$ fractional D3-brane stack in (\ref{cps}) can end on the $i^{th}$ stack of 5-branes. We also note that, since a complete $\IZ_k$ orbit of fractional D3-branes (a regular D3-brane) behaves as a D3-brane in the covering space, it inherits the interpretation of boundary conditions of the parent 4d $\NN=4$ theories, namely the quantities $K_a$, $L_b$. Hence, they are naturally encoded in the dual geometries, which we consider next.

\medskip

The gravity duals of the 4d $\NN=2$ theories coupled to the BCFT are just an orbifold of those considered in section \ref{sec:case-of-one}. Recall that we have a warped AdS$_4\times \IS_1^2\times\IS_2^2$ varying over the Riemann surface $\Sigma$ parametrized by a quadrant in the $w$-plane. The fibration preserves an $SO(3)\times SO(3)$ symmetry, corresponding to the rotation of the 3-planes 456 and 789 in the brane picture. This means that the $\IZ_k$ action (\ref{zk-generator}) is in a subgroup of $SO(3)_1\times SO(3)_2$. 
Let us introduce angular coordinates in the 3-planes 456 and 789 to parametrize the  $\IS^2$'s as follows,
\beqa
&x_4= R_1 \cos\theta_1 \sin\varphi_1\quad , \quad &x_7= R_2 \cos\theta_2 \sin\varphi_2\nonumber \\
& x_5=R_1 \cos\theta_1\cos\varphi_1\quad,\quad & x_8= R_2\cos\theta_2\cos\varphi_2\nonumber\\
&x_6=R_1\sin\theta_1 \quad,\quad & x_9=R_2\sin\theta_2
\label{spherical-coords}
\eeqa
The generator $\Theta$ of $\IZ_k$ acts as
\beqa
\Theta:\; \varphi_1\rightarrow e^{2\pi i/k}\varphi_1 \quad ,\quad  \varphi_2\rightarrow e^{-2\pi i/k}\varphi_2\, ,
\eeqa
leaving $\theta_1,\theta_2$ invariant, as well as $w$ on the base $\Sigma$. 

At a generic position in $w$, the above action on $\IS_1^2\times \IS_2^2$ has 4 fixed points, locally of the form $\IC^2/\IZ_k$, with coordinates at the poles of the spheres. Hence, one might think that there are four copies of the corresponding twisted sector fields, propagating over $w$. This replication of fields would disagree with those in the AdS$_5\times\IS^5/\IZ_k$ dual; it would also be puzzling because $w$ parametrizes just a quadrant $\Sigma$, so  the geometry on which these twisted fields propagate would have boundaries. Both problems are related, and are happily solved at once if we notice that the boundaries in $\Sigma$ correspond to points where some $\IS^2$ shrinks to zero, which leads to a reduced number of fixed points. In other words, the fixed point set of the $\IZ_k$ action is a four-sheeted cover of $\Sigma$, with different sheets joining at the boundaries of $\Sigma$. One can check that the complete fixed set is topologically a complex plane (an orientable non-compact Riemann surface with no handles and no boundaries) as follows: 

Consider the polar coordinates $w=re^{i\varphi}$ and consider the $\IS^5$ given by the fibration of $\IS_1^2\times \IS_2^2$ over $\varphi$. At a generic point in $\varphi$, there are four fixed points of the $\IZ_k$ action on $\IS_1^2\times \IS_2^2$, which we denote $P_{ij}$, with $i,j=1,2$ labeling the two poles on each $\IS^2$. However, at one of the endpoints of $\varphi$ one of the 2-spheres e.g. $\IS_1^2$ shrinks to zero and the four points collapse pairwise into two, $P_{11}=P_{21}$, $P_{12}=P_{22}$; at the other endpoint of $\varphi$, $\IS_2^2$ shrinks to zero size and the four points collapse to two in a different pattern $P_{11}=P_{12}$, $P_{21}=P_{22}$. It is easy to see that in the whole fibration over $\varphi$, the 4-sheeted fixed point set  describes a single $\IS^1$ with no boundary. Adding now the radial coordinate builds a real cone over this $\IS^1$, showing the complete fixed point set is a complex plane, as announced. Incidentally, note that the  above $\IS^1\subset \IS^5$ is precisely the geometry of the fixed point set in the AdS$_5\times \IS^5/\IZ_k$ gravity dual of the bulk $\NN=2$ theory; this is as should be, since the 10d ETW geometry reproduces precisely such gravity dual in the large $r$ regime.

The fact that the fixed point set has no non-trivial topology in the ETW brane region suggests that there are no localized modes of the bulk twisted sector fields in the KR brane for these AdS$_5$ geometries. Hence, they are adequately described by the effective action description in section \ref{sec:eft}, by simply reinterpreting $\phi_+$ as the breathing mode of $\IS^5/\IZ_k$.

\subsection{Boundaries from Brane Box construction}
\label{sec:branebox}

It is known that 4d $\NN=2$ quiver theories admit a brane construction in terms of a configuration of D4-branes spanning 0123 and suspended in a direction 6 between sets of NS5 branes spanning 012345 \cite{Witten:1997sc}. Linear quiver theories are obtained by having non-compact direction 6, while cyclic quivers are obtained when the direction 6 parametrizes an $\IS^1$. The latter correspond to the theories of D3-branes at $\IC^2/\IZ_k$ singularities, and both setups are in fact related \cite{Karch:1998yv} by a T-duality along 6, which turns the suspended D4- into fractional D3-branes, and NS5-branes into the orbifold geometry. 

In this section we show that the T-duality continues to hold in the presence of the 5-brane configurations which define boundary conditions for the 4d $\NN=2$ theories on D3-branes at $\IC^2/\IZ_k$ singularities. In other words, we describe how to introduce sets of 5-branes in the above Hanany-Witten NS5/D4-brane configurations so as to introduce boundary conditions.

Consider the following sets of branes 
\begin{center}
\begin{tabular}{ccccccccccc}
D4 & 0 & 1 & 2 & 3 & $\times$ & $\times$ & 6 & $\times$ & $\times$ & $\times$ \\
NS5 & 0 & 1 & 2 & 3 & 4 & 5& $\times$ & $\times$ & $\times$ & $\times$ \\
NS5' & 0 & 1 & 2 & $\times$ & 4 & $\times$ & 6 & 7 & $\times$ & $\times$ \\
\end{tabular}
\end{center}
where a cross indicates the brane is localized in that direction, and NS5'-branes are primed to emphasize their different orientation with respect to NS5-branes.

\begin{figure}[htb]
\begin{center}
\includegraphics[scale=.25]{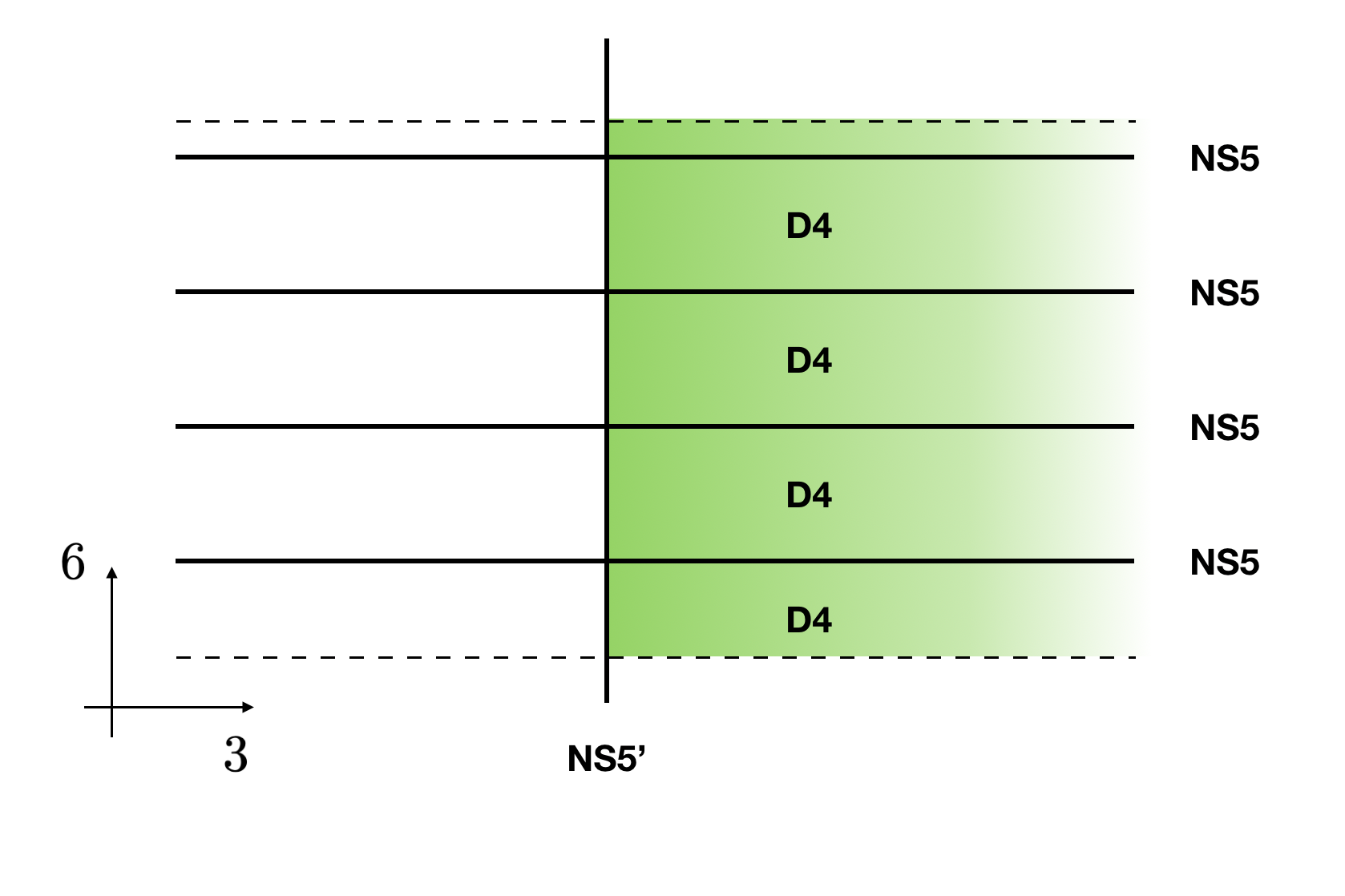}
\caption{\small Configuration of D4-branes bounded by NS5- and NS5'-branes. The direction 6 is compact by the identification of the dashed lines. The direction 3 is part of the dimensions of the 4d $\NN=2$ quiver gauge theory  with $k=4$ defined by the D4-branes suspended among the NS5-branes in the direction 6; the NS5'-branes define the boundary in the direction 3.}
\label{fig:bb}
\end{center}
\end{figure}

It is easy to check that the system preserves 4 supercharges. As above, the D4-branes are suspended in 6 between different NS5-branes, and we make the direction 6 periodic. In addition, we introduce a single stack of NS5'-branes and we specify that the D4-brane end on the NS5'-branes, so that they are semi-infinite in the direction 3, see Figure \ref{fig:bb}. The 4d theory, arising from the NS5- and D4-brane configurations, is 4d $\NN=2$, but the introduction of NS5'- branes breaks the 4d Poincar\'e invariance and breaks the supersymmetry by 1/2. This is a confirmation of the symmetry preserved by the boundary conditions we have discussed in the previous sections. 

Note that this is a D4-brane version of the D5-brane brane box configurations introduced in \cite{Hanany:1997tb}, with the NS' brane bounding semi-infinite boxes, filled with D4-branes on one side and empty on the other, to reproduce the bounded dimension of the 4d gauge theory. We also note that the T-duality along 6, relating the NS5/D4-systems to D3-branes at $\IC^2/\IZ_k$, also maps the NS5'-branes into the same kind of NS5-branes in the previous section.

We may also consider the brane box version of the D5-branes defining boundaries in the picture of D3-branes at $\IC^2/\IZ_k$. They are easily identified as D6-branes with the following orientation 
\begin{center}
\begin{tabular}{ccccccccccc}
D6' & 0 & 1 & 2 & $\times$ & $\times$ & 5 & 6 & $\times$ & 8 & 9 \\
\end{tabular}
\end{center}
(where they are primed in analogy with the NS5'-branes).
These can be introduced in the earlier configuration of NS5/D4-branes and NS5'-branes, still preserving 4 supersymmetries. Since the D6'-branes are localized in the direction 3, they admit D4-branes ending on them and hence define boundary conditions

Hence, the set of NS5'- and D6'-branes provide a set of boundary conditions, specified by the partitioning of the set of $N$ gauge D4-branes (for each quiver node) into sets ending on different boundary branes. As in the previous discussions, the interpretation of the quantities $K_a$, $L_b$ remains as describing the numbers of (regular) D3-branes ending on a given 5-brane with boundary conditions associated to some $SU(2)$ representation, in the parent theory.

\section{ETW branes in gravity duals of $\NN=3$ S-fold theories}
\label{sec:sfolds}

In this section we extend the logic of the previous section, generalizing to quotients including non-trivial duality relations. We focus on the particular case of S-folds \cite{Garcia-Etxebarria:2015wns}.

\subsection{The 4d $\NN=3$ theories on D3-branes}
\label{sec:quiverfold}

Let us go back to the configuration of D3-branes in flat space and perform an orbifold quotient of the kind considered in \cite{Garcia-Etxebarria:2015wns}. Namely, we combine coordinates as
\beqa
z_1=x_4+ix_7\quad ,\quad z_2=x_5+ix_8\quad ,\quad z_3=x_6+ix_9\, ,
\eeqa
and we also introduce the coordinate $z_4$ for the F-theory $\IT^2$, 
and mod by
\beqa
\Theta: z_k\to e^{2\pi i v_k}z_k\, ,
\eeqa
with 
\beqa
v=\big(\,\frac 1k,\frac 1k,\frac 1k,\frac 1k\,\big)\, .
\eeqa
This is the action in \cite{Garcia-Etxebarria:2015wns}, up to an irrelevant change in the choice of complex structure. The allowed values are $k=2,3,4,6$, so as to have a consistent crystallographic action on the F-theory $\IT^2$.

We are interested in considering S-fold actions of this kind leaving invariant the configuration of NS5- and D5-branes providing boundary conditions for the D3-branes as in section \ref{sec:holobound}.
The only possible choices are $k=2,4$. The simple choice is $k=2$, which as explained in \cite{Garcia-Etxebarria:2015wns} corresponds to introducing an O3-plane on top of the D3-branes, and actually preserves 4d $\NN=4$. The O3-plane maps the NS5-branes to themselves and the D5-branes to themselves, so their structure of numbers of stacks and multiplicities are independent. Also, the string coupling is arbitrary, as the action of $\Theta$ on $\tau$ is trivial. The gravity dual will be discussed in the next section.

The only genuinely $\NN=3$ case hence corresponds to $k=4$, generated by 
\beqa
\Theta: (z_1,z_2,z_3,z_4)\to  (i z_1,iz_2,iz_3,iz_4)\, .
\eeqa
In terms of the cartesian coordinates we have
\beqa
\Theta: &&(x_4,x_5,x_6)\to (x_7,x_8,x_9)\nonumber\\
&&(x_7,x_8,x_9)\to (-x_4,-x_5,-x_6)
\label{z4action-1}
\eeqa
and a simultaneous $SL(2,\IZ)$ action by the monodromy {\footnotesize
$\begin{pmatrix}
0&1\cr-1&0
\end{pmatrix}
\label{mono}
$}, 
which acts on $\tau$ as $\tau\to -1/\tau$, so it is the $S$ generator of $SL(2,\IZ)$. Hence, in this 
$k=4$ case, the NS5-branes are mapped to the D5-branes and viceversa\footnote{We note that the element $\Theta^2$ in the S-fold group corresponds to the above O3-plane $\IZ_2$ action. So acting twice with $\Theta$ brings each 5-brane to itself, up to this orientifold action.}, due to the simultaneous rotation of the 3-planes they span and the $SL(2,\IZ)$ action on them. This implies that the structure of numbers of stacks and multiplicities of must be identical for NS5- and D5-branes. Also, the solution necessarily contains regimes of strong coupling, as in the parent S-fold theory.

\subsection{The ETW configurations in the gravity duals}
\label{sec:etw-sfold}

In this section we sketch some properties of the ETW configurations gravitational dual to the above BCFT configurations for the above 4d $\NN=3$ S-fold theories.

\subsubsection*{The $k=2$ case}

We start with the $k=2$ case. In the absence of boundaries, the gravity dual of the D3-brane theory is AdS$_5\times {\bf RP}_5$ \cite{Witten:1998xy}, where ${\bf RP}_5$ is the quotient of $\IS^5$ by the orientifold action $\Omega(-1)^{F_L}{\cal R}$, with ${\cal R}$ sending any point to its antipodes in $\IS^5$. Writing the $\IS^5$ in the $\IR^6$ transverse to the D3-branes as
\beqa
x_4^2+x_5^2+x_6^2+x_7^2+x_8^2+x_9^2=R^2
\label{s5}
\eeqa
the action is simply inherited from ${\cal R}: x^i\to -x^i$. Note that, since the origin $x_i=0$ is the only fixed point, the $\IZ_2$ is acting freely on the $\IS^5$ and there are no singularities. 

In the presence of 5-branes defining boundaries, the ETW configuration is obtained from the ETW configuration of section \ref{sec:case-of-one} by quotienting by the orientifold action $\Omega(-1)^{F_L}{\cal R}$. The geometric action ${\cal R}$ acts on the $\IS^2\times \IS^2$ fibration over $\Sigma$ as follows: it leaves $w$ invariant, while it acts on each $\IS^2$ sending any point to its antipode in $\IS^2$. This just follows from the fact that the $\IS^2$'s are given by the angular piece of the $\IR^3$'s, on which ${\cal R}$ acts by flipping the 3 coordinates. It is easy to see that this agrees with the action of ${\cal R}$ in the asymptotic AdS$_5\times \IS^5$ region.

An interesting point is that, although the $\IZ_2$ quotient has no fixed points on the $\IS^5$, there is a fixed point at $w=0$, because the $\IS^5$ collapses (and so do the $\IS^2$'s).  In the F-theory description, this is locally described by a $\IC^4/\IZ_2$ singularity. This is a familiar example of terminal singularity, i.e. one not admitting a crepant resolution (i.e. there is no blow-up preserving the Calabi-Yau property). This nicely agrees with the fact that, in the perturbative description, the singularity corresponds to an O3-plane at $w=0$, and O3-planes carry no localized degrees of freedom.

\subsubsection*{The $k=4$ case}

Consider now the $k=4$ case. In the absence of boundaries for the 4d $\NN=3$ S-fold theory, the gravity dual is given by an F-theory fibration over AdS$_5\times \IS^5/\IZ_4$. The action of the $\IZ_4$ is obtained from (\ref{z4action-1}) by simply cutting out the $\IS^5$ shell in $\IR^6$, c.f. (\ref{s5}). Again, since the origin $x_i=0$ is the only fixed point under (\ref{z4action-1}), the $\IZ_4$ is acting freely on the $\IS^5$, and there are no fixed points.

As explained, in the presence of 5-branes defining boundaries, the ETW configuration is obtained by starting with a parent ETW configuration of section \ref{sec:case-of-one} with symmetric distributions of NS- and D5-brane poles. Namely, we take $n=m$, and symmetric poles $k_a=l_a$ and coefficients $d_a={\tilde d}_a$. Then, also using $w=-x+iy$ in (\ref{the-hs}), we have
\beqa
&&h_1=\;4y+2\sum_{a=1}^n d_a \log\left(\frac{x^2+(y+k_a)^2}{x^2+(y-k_a)^2}\right)\nonumber\\
&&h_2=4x+2\sum_{a=1}^n d_a \log\left(\frac{(x+k_a)^2+y^2}{(x-k_a)^2+y^2}\right)\, ,
\label{the-hs-n3}
\eeqa
We then have to perform the $\IZ_4$ quotient, making sure we recover the right behaviour in the asymptotic AdS$_5\times \IS^5$ region, namely (\ref{z4action-1}) (plus the $SL(2,\IZ)$ monodromy). Let us split (\ref{s5}) in two equations, to display more explicitly the phase $\varphi$ of the base coordinate $w=re^{i\varphi}$
\beqa
&&x_4^2+x_5^2+x_6^2=R^2 \cos^2\varphi \equiv R_1^2(\varphi)\nonumber\\
&& x_7^2+x_8^2+x_9^2=R^2 \sin^2\varphi\equiv R_2^2(\varphi)
\eeqa
The action (\ref{z4action-1}) of $\Theta$ requires exchanging these two equations, namely 
\beqa
\Theta: \varphi\to \frac{3\pi}2-\varphi
\label{actionz4-2}
\eeqa
so that the values $\varphi=\pi/2$ and $\varphi=\pi$ are exchanged, and so are the NS5- and D5-brane poles. This corresponds to the exchange $x\leftrightarrow y$ in (\ref{the-hs-n3}), clearly a symmetry.
By propagating the exchange $h_1\leftrightarrow h_2$ through (\ref{wnn}), (\ref{dilaton}), we see that it corresponds to $\phi\to -\phi$, which agrees with the $SL(2,\IZ)$ action.

Although the action of $\Theta$ on $\Sigma$ is just a $\IZ_2$ flip, we should not forget that there is also the action on the $\IS^2$'s. To determine this, let us introduce the coordinates $\theta_1$, $\varphi_1$ and $\theta_2$, $\varphi_2$ as in (\ref{spherical-coords}). Noticing that the $\IS^2$'s are the angular manifolds in the 3-planes 456 and 789, the action (\ref{z4action-1}) corresponds to
\beqa
\Theta: & \theta_1\to \theta_2 \quad, \quad & \theta_2\to -\theta_1 \nonumber\\
& \varphi_1\to\varphi_2\quad,\quad & \varphi_2\to \varphi_1+\pi
\label{actionz4-3}
\eeqa
This exchanges the two $\IS^2$, but with a $\IZ_4$ action: a point of $\IS_1^2$ goes to the corresponding point in $\IS_2^2$, and a point of $\IS_2^2$ goes the antipodes of the corresponding point in $\IS_1^2$. This also shows that there are no fixed points of this action on $\IS_1^2\times \IS_2^2$. 
The action of $\Theta$ on the ETW configuration is thus the combination of (\ref{actionz4-2}) and (\ref{actionz4-3}).

An interesting feature, similar to the $k=2$ case, is that, although there are no fixed points in the asymptotic  AdS$_5\times \IS^5/\IZ_4$ case,  there is a fixed point in the 10d ETW configuration at the origin $w=0$ in $\Sigma$, where the $\IS^2$'s shrink to zero size and (\ref{actionz4-3}) is not freely acting. 

The singular point is locally given by F-theory on $\IC^4/\IZ_4$, so it is strongly coupled and hence non-trivial to analyze. We are not aware of a discussion of this singularity in the F-theory literature. Anyway, in analogy with the $\IZ_2$ case in the previous section, the $\IC^4/\IZ_4$ with orbifold vector $v=(1/4,1/4,1/4,1/4)$ is terminal \cite{Morrison:1984}. This also agrees with the absence of massless twisted sector states of the type IIA perturbative orbifold \cite{Font:2004et}, which relates to the F-theory setup via dualities. In brief, the S-fold fixed point seems to have no localized degrees of freedom, in agreement with its interpretation as a non-perturbative generalization of O3-planes. Hence we expect no additional modes localized on the ETW brane from the appearance of this singular point.

The absence of localized twisted modes in the KR brane for these AdS$_5$ geometries implies that they are adequately described by the effective action description in section \ref{sec:eft}, by simply reinterpreting $\phi_+$ as the breathing mode of the quotiented $\IS^5$.

\section{Conclusions}
\label{sec:conclusions}

In this work we have studied the 10d $SO(3)\times SO(3)$ preserving solutions defining the gravity dual of 4d $\NN=4$ SYM with boundary defined by a 3d $\NN=4$ BCFT (and quotientes thereof), emphasizing their interpretation as cobordisms to nothing for the corresponding AdS$_5\times\IS^5$ geometries. We have provided a description in terms of dynamical cobordisms in the 5d theory after reduction on the $\IS^5$, and shown that they describe a generalization of the AdS$_4$ KR branes. Our 5d solutions display a smoothing out of the KR cusp into a bump, which still leads to localization of gravity. The smoothing out is due to the presence of additional scalars, which encode the varying internal geometry; in particular, there is a bulk $SO(6)$ invariant breathing mode which describes the shrinking of $\IS^5$ and the end of spacetime, and there are localized modes which encode the $SO(6)\to SO(3)\times SO(3)$ breaking due to the boundary conditions. We have clarified the conditions of the 10 solution under which such gravity localization arises, and we have introduced a new scaling limit of the 10d solution which isolates the properties of the ETW brane. Finally, we have extended the discussion to several orbifold and S-fold quotients, whose ETW branes are inherited from those of the parent theory; we have shown that, despite the presence of genuinely new singular points in the quotient, they lead to no new localized degrees of freedom.

There are several interesting directions to be pursued:

$\bullet$ It would be interesting to study gravity duals of 4d $\NN=1$ theories with boundaries. The naive extension of our construction for 4d $\NN=2$ theories, namely using 4d $\NN=1$ theories on D3-branes at $\IC^3/\IZ_k$ singularities \cite{Kachru:1998ys,Lawrence:1998ja,Ibanez:1998xn,Hanany:1998it,Aldazabal:2000sa}, do not work: the orbifold does not preserve the geometry of the NS5- and D5-branes of the parent theory. This difficulty is linked to the fact that the 4d $\NN=1$ orbifold theories are generically chiral, and so to the difficulties to introduce boundaries in chiral theories.

$\bullet$ Even in the 4d $\NN=4$ theory, it is possible to introduce boundaries preserving less (super)symmetry, by simply using NS5- and D5-branes with different orientations. A simple proposal, inspired by type IIA brane setups \cite{Elitzur:1997fh,Barbon:1997zu,Elitzur:1997fh} is to use NS5 branes along 012456 and 012678, and D5-branes along 012789 and 012457. There is no known 10d solution for the corresponding gravity dual, but it may be possible to prescribe an approximate 5d effective theory, based on the our same $SO(6)$ invariant sector, but a richer set of ETW brane localized modes to account for the further breaking of symmetries.

$\bullet$ It would be interesting to exploit the tools of AdS/BCFT to gain further understanding of the dynamics of the ETW localized modes and their interplay with AdS$_5$ bulk modes.

$\bullet$ Finally, it would be interesting to extend the present analysis to other holographic setups, such as AdS$_4\times\IS^7$ and AdS$_7\times\IS^4$, possibly based on the solutions discussed in \cite{DHoker:2008lup,DHoker:2008rje,DHoker:2009lky,DHoker:2009wlx}.

We hope to come back to these and other questions in the future.

\medskip


%
\section*{Acknowledgments}

We are pleased to thank Matilda Delgado, Miguel Montero and Irene Valenzuela for useful discussions originating this project, and Roberta Angius, Jos\'e Calder\'on-Infante, Anamaria Font, I\~naki Garcia-Etxebarria, Luis E. Ib\'a\~nez and Fernando Marchesano for useful discussions.  
This work is supported through the grants CEX2020-001007-S and PID2021-123017NB-I00, funded by MCIN/AEI/10.13039/501100011033 and by ERDF A way of making Europe. The work by J. H. is supported by the FPU grant FPU20/01495 from the Spanish Ministry of Education and Universities.

\newpage

\bibliographystyle{JHEP}
\bibliography{mybib}

\end{document}